\documentclass[a4paper,12pt]{article}
\usepackage{jheppub}

\numberwithin{equation}{section}

\newcommand{\We}{\mathcal{W}_{\infty}^{e}}
\newcommand{\hse}{\mathrm{hs}^e[\mu]}
\newcommand{\WeB}{\mathcal{WB}_{\infty}}
\newcommand{\WeC}{\mathcal{WC}_{\infty}}

\newcommand{\be}{\begin{equation}}
\newcommand{\ee}{\end{equation}}

\newcommand{\hs}[1]{\mbox{hs$[#1]$}}

\newcommand{\W}{\mathcal{W}}

\numberwithin{equation}{section}

\def\be{\begin{equation}}
\def\ee{\end{equation}}

\title{Even spin minimal model holography}

\author{Constantin Candu,}
\author{Matthias R.\ Gaberdiel,}
\author{Maximilian Kelm}
\author{and Carl~Vollenweider}
\affiliation{Institut f\"ur Theoretische Physik, ETH Z\"urich, \\
CH-8093 Z\"urich, Switzerland}
\emailAdd{canduc@itp.phys.ethz.ch}
\emailAdd{gaberdiel@itp.phys.ethz.ch}
\emailAdd{mkelm@itp.phys.ethz.ch}
\emailAdd{carlv@itp.phys.ethz.ch}

\abstract{The even spin $\W^e_\infty$ algebra that is generated by the stress energy tensor together with
one Virasoro primary field for every even spin $s\geq 4$
is analysed systematically by studying the constraints coming from the Jacobi identities. It is found that the algebra 
is characterised, in addition to the central charge, by one free parameter that can be identified with the self-coupling
constant of the spin $4$ field. We show that $\W^e_\infty$ can be thought of as the quantisation of the asymptotic
symmetry algebra of the even higher spin theory on AdS$_3$. On the other hand, $\W^e_\infty$ is also 
quantum equivalent to the $\mathfrak{so}(N)$ coset algebras, and thus our result establishes
an important aspect of the even spin minimal model holography conjecture. The quantum equivalence holds actually
at finite central charge, and hence opens the way towards understanding the duality beyond the leading 't~Hooft limit.
}

\allowdisplaybreaks

\begin{document}
\maketitle


\section{Introduction}


Higher spin holography is a simplified example of Maldacena's 
celebrated anti de Sitter / conformal field theory
(AdS/CFT) correspondence \cite{Maldacena:1997re}, promising 
interesting insights into the mechanisms underlying the duality.
The bulk theory is a Vasiliev higher spin gauge theory on AdS \cite{Vasiliev:1989re}, 
which has been argued to be dual to the singlet sector of vector-like CFTs
\cite{Sundborg:2000wp,Witten,Mikhailov:2002bp,Sezgin:2002rt}. 
Since both theories are in a perturbative regime, the duality can be checked and understood quite
explicitly. Although the theories need not be supersymmetric,
their symmetry algebras are large enough to constrain the theories considerably
\cite{Maldacena:2011jn,Maldacena:2012sf}.

A concrete realisation of this idea was proposed some time ago by Klebanov \& Polyakov
\cite{Klebanov:2002ja} (and shortly later generalised further in \cite{Sezgin:2003pt}). It relates 
a higher spin theory on AdS$_4$ to  the singlet sector of the $3$-dimensional ${\rm O}(N)$ vector model
in the large $N$ limit, for a recent review see
e.\,g.\  \cite{Giombi:2012ms}. More recently, the lower dimensional 
version of this duality, connecting a 
higher spin theory on AdS$_3$ to a 2-dimensional CFT, was conjectured in
\cite{Gaberdiel:2010pz} (see \cite{Gaberdiel:2012uj} for a review). One advantage
of these low dimensional theories is that they are comparatively well understood while
avoiding no-go theorems of the Coleman-Mandula type as in \cite{Maldacena:2011jn}.
The  2d CFTs are large $N$ limits of minimal models about which much is known, while the
bulk theory can be formulated as a Chern-Simons gauge theory based on
$\hs{\mu}$ \cite{Prokushkin:1998bq,Prokushkin:1998vn} (see also \cite{Vasiliev:1999ba}).
Here $\hs{\mu}$ is an infinite-dimensional Lie algebra extending the $\mathfrak{sl}(2)$
gauge algebra
of pure gravity on AdS$_3$. The resulting theory generically contains an
infinite tower of higher spin gauge fields. However, for $\mu=N\in\mathbb{N}$, $\hs{\mu}$ reduces to 
$\mathfrak{su}(N)$, and the higher spin theory truncates to a theory containing
higher spin gauge fields of spin $s=2,\ldots,N$. In addition 
to the gauge fields, the theory also contains 
(one or two) complex scalar fields. 

The classical asymptotic symmetry algebra of these higher spin theories was determined  in
\cite{Henneaux:2010xg,Campoleoni:2010zq,Gaberdiel:2011wb}, 
following the old analysis of Brown \& Henneaux \cite{Brown:1986nw}.
It is described by the non-linear Poisson algebra 
$\W_{\infty}^{\rm cl}[\mu]$, where `non-linear' means that the Poisson
brackets can in general only be expressed in terms of polynomials of the generators, see 
\cite{Bouwknegt:1992wg} for a review on $\W$ algebras. 
This asymptotic symmetry algebra can be realised as a classical Drinfel'd-Sokolov reduction of
$\hs{\mu}$ \cite{Khesin:1994ey,Campoleoni:2011hg}.
Furthermore, it was argued to agree \cite{Gaberdiel:2011wb} with the (semiclassical) 't~Hooft limit
of the $\W$ algebras of the coset models $\mathcal{W}_{N,k}$, provided one 
identifies $\mu$ with the 't~Hooft parameter $\lambda=\frac{N}{N+k}$ that is held constant in the large $N,k$ limit.

In order to make sense of a similar statement for finite $N$ and $k$, one must first define
the quantisation of $\W_{\infty}^{\rm cl}[\mu]$. This is non-trivial 
since the classical algebra is non-linear,
and hence a naive replacement of Poisson brackets by commutators 
does not lead to a consistent Lie algebra
(satisfying the Jacobi identity). It was recently shown 
in \cite{Gaberdiel:2012ku} how this problem can be 
overcome (by adjusting the structure constants), and how the resulting quantum algebra 
$\W_{\infty}[\mu]$ can be defined for arbitrary central charge; it  can therefore be interpreted as the
quantum Drinfel'd-Sokolov reduction of $\hs{\mu}$. It was 
furthermore shown in \cite{Gaberdiel:2012ku} that
$\W_{\infty}[\mu]$ exhibits an intriguing triality relation that implies, among other things, that
we have the equivalence of quantum algebras 
$\W_{N,k} \cong  \W_{\infty}[\mu=\lambda]$ at the central charge $c=c_{N,k}$ of the minimal models. This 
relationship can be interpreted as a generalisation of the level-rank dualities of coset algebras
proposed in  \cite{Kuniba:1990zh,Altschuler:1990th}.

\smallskip

In this paper we will extend these results to the 
higher spin theory on AdS$_3$ that contains only gauge fields 
of even spin. This is the natural analogue of the
Klebanov \& Polyakov proposal, which involves the smallest (or minimal) higher spin theory 
on AdS$_4$. For the case of AdS$_3$, the gauge symmetry 
can be described by a Chern-Simons theory based on a 
suitable subalgebra of $\hs{\mu}$, and it was argued in 
\cite{Ahn:2011pv,Gaberdiel:2011nt} (see also \cite{Ahn:2012gw}) that it should be 
dual to the ${\rm SO}(N)$ coset theories of the form
\begin{equation*}
\frac{\mathfrak{so}(N)_k \oplus \mathfrak{so}(N)_1}{\mathfrak{so}(N)_{k+1}} \ .
\end{equation*}
While the (classical) asymptotic symmetry algebra of the bulk theory has not 
yet been determined explicitly, it is clear that it
will be described by a classical $\W$ algebra 
that is generated by one field for every even spin $s=2,4,\ldots$\, . 
One expects on general grounds that it will
be non-linear, and hence the quantisation will exhibit the same subtleties as 
described above. As a consequence,
it is actually simpler to approach this problem by constructing directly 
the most general {\em quantum} $\W$ algebra 
$\We[\mu]$ with this spin content. 
As in the case of $\W_{\infty}[\mu]$, one finds that the successive Jacobi identities
fix the structure constants of all commutators in terms of a single 
parameter $\gamma$, as well as the central charge. 
For a suitable identification of $\gamma$ and $\mu$, we can then think of these 
algebras as the quantum Drinfel'd-Sokolov
reduction of some subalgebra of $\hs{\mu}$, which turns out to be 
the $\hse$  algebra of \cite{Gaberdiel:2011nt}. However, compared to the $\W_{\infty}[\mu]$ analysis of 
\cite{Gaberdiel:2012ku}, there is an unexpected subtlety in that there are two 
natural ways in which one may identify 
$\gamma$ and $\mu$ at finite $c$ --- the two identifications agree in the 
quasiclassical $c\rightarrow \infty$ limit, but
differ in their $1/c$ corrections. This
reflects the fact that $\hse$  truncates for $\mu=N$ to either $\mathfrak{sp}(N)$ (if $N$ is even), or 
$\mathfrak{so}(N)$ (if $N$ is odd), and that the Drinfel'd-Sokolov 
reduction of these non-simply-laced algebras 
are Langlands dual (rather than equivalent).

Since the quantum algebra $\We[\mu]$ is the most general $\W$ algebra 
with the given spin content, we can
also identify the $\mathfrak{so}$ and $\mathfrak{sp}$ cosets 
(or rather their orbifolds) with these algebras. In this way we obtain 
again non-trivial identifications between quantum $\W_{\infty}^{e}$ algebras 
that explain and refine the holographic conjectures of 
\cite{Ahn:2011pv} and \cite{Gaberdiel:2011nt}, see eqs.~(\ref{eq:dbhol}) and (\ref{eq:dchol}) below. 
Furthermore, there are again non-trivial quantum equivalences between
the algebras for different values of $\mu$, which can be interpreted in terms of 
level-rank dualities of $\mathfrak{so}$ coset models that do not seem to have been noticed before,
see eq.~\eqref{eq:mm_lrd}. 
 
\medskip

The paper is organised as follows. In section~2 we construct 
the most general quantum $\We$ algebra, and explain
how the different structure constants can be determined recursively 
from the Jacobi identities. We also consider
various truncations to finitely generated algebras that have 
been studied in the literature before (see section~2.2),
and explain that the wedge algebra of $\We$ is indeed the $\hse$ algebra of \cite{Gaberdiel:2011nt}.
Section~3 is devoted towards identifying $\We[\mu]$ as a Drinfel'd-Sokolov reduction of $\hse$. As in 
\cite{Gaberdiel:2012ku} the relation between the two algebras can be most easily an\-a\-lysed by studying 
some simple representations of the two algebras. It turns out 
that there is no canonical identification, but
rather two separate choices that we denote by ${\cal WB}_{\infty}[\mu]$ and ${\cal WC}_{\infty}[\mu]$,
respectively; this nomenclature reflects the origin of this ambiguity, 
namely that $\hse$ truncates to either
$C_n=\mathfrak{sp}(2n)$ or $B_n = \mathfrak{so}(2n+1)$, depending on whether $\mu=N$ is even or odd.

In section~4 we apply these results to the actual higher spin holography. 
In particular, we show that the (subalgebras of the) 
$\mathfrak{so}$ cosets fit into this framework, and hence deduce the precise relation between 
${\cal WB}_{\infty}[\mu]$ or ${\cal WC}_{\infty}[\mu]$ and the $\mathfrak{so}$ coset algebras
at finite $N$ and $k$. We comment on the fact that the matching of 
the partition functions requires that we consider
a non-diagonal modular invariant with respect to the orbifold subalgebra of the $\mathfrak{so}$ cosets 
(see section~4.6). We also explain that the non-trivial identifications among the $\We$ algebras imply a 
level-rank duality for the $\mathfrak{so}$ cosets themselves, and that also the cosets based on 
$\mathfrak{sp}(2n)$ and $\mathfrak{osp}(1|2n)$ can be brought into the fold.
Finally, we show that, as in \cite{Gaberdiel:2012ku}, only one of the two real scalars 
in the bulk theory should be thought of as being perturbative. 

Section~5 contains our conclusions as well
as possible directions for future work. There are two appendices, 
where some of the more technical material
has been collected.


\section{The even spin algebra}

\subsection{Construction}
\label{sec:W}

In this section we analyse the most general ${\cal W}_{\infty}$ algebra $\We$ that is generated
by the stress energy tensor $L$ and one Virasoro primary field $W^s$ for each 
even spin $s=4,6,\ldots$\;. As we shall see, the construction
allows for one free parameter in addition to the central charge.

The strategy of our analysis is as follows. First we make the most general ansatz for
the OPEs of the  generating fields $W^s$ with each other. In a second step we then 
impose the constraints that
come from solving the various Jacobi identities. Actually, instead of working directly in terms
of modes and Jacobi identities, it is more convenient to do this analysis on the level of the OPEs. Then 
the relevant condition is that the OPEs are associative. 


\subsubsection{Ansatz for OPEs}


We know on general grounds that the 
conformal symmetry, i.\,e.\ the associativity of the OPEs involving the stress energy tensor $L$,
fixes the coefficients of the Virasoro descendant fields in the OPEs in  terms of the Virasoro primary 
fields. In order to make the most general ansatz we therefore only have to introduce
free parameters for the coupling to the Virasoro primary fields. Thus we need to know
how many Virasoro primary fields the algebra $\We$ contains. 
This can be determined by decomposing the vacuum character of $\We$
\begin{equation}\label{eq:ddef}
\chi_\infty (q)=\mathrm{Tr}_0 q^{L_0} = \prod_{s \in 2 \mathbb{N}} \prod_{n=s}^\infty \frac{1}{1-q^n}= \chi_0(q)+\sum_{h=4}^\infty d(h) \chi_h(q) \ ,
\end{equation}
in terms of  the Virasoro characters corresponding to the vacuum representation $\chi_0 (q)$, 
and to a highest weight representation with conformal dimension $h$ 
\be\label{chih}
\chi_0(q) = \prod_{n=2}^{\infty} \frac{1}{1-q^n} \ , \qquad
\chi_h (q) = q^{h}\prod_{n=1}^\infty \frac{1}{1-q^n} \ .
\ee
Note that since we are working at a generic central charge, there are no Virasoro null-vectors. 
The coefficients $d(h)$ in (\ref{eq:ddef}) are then the number of Virasoro primary fields of 
conformal dimension $h$. Their generating function equals 
\begin{equation}\label{eq:Palg}
P(q)=\sum_{h=4}^\infty d(h) q^h = (1-q)(\chi_{\mathrm{HS}}(q)-1)=q^4+q^6+2 q^8+3q^{10}+q^{11}+6q^{12}
+\cdots\ ,
\end{equation}
where $\chi_{\mathrm{HS}}(q)=\chi_\infty(q)/\chi_0(q)$
denotes the contribution of the higher spin fields to the character $\chi_\infty$.

\noindent 
The most general ansatz for the OPEs  is then
\begin{equation}\label{eq:wope}
\begin{split}
W^4 \star W^4 &\sim c_{44}^6 W^6 + c_{44}^4 W^4 + n_4 I\ ,\\
W^4 \star W^6 &\sim c_{46}^{8} W^8 + a_{46}^8 A^8 + c_{46}^6 W^6 + c_{46}^4 W^4\ ,\\ 
W^4 \star W^8 &\sim a_{48}^{11} A^{11} + c_{48}^{10} W^{10} + a_{48}^{10,1} A^{10,1} +  a_{48}^{10,2} A^{10,2} + c_{48}^8 W^8 + a_{48}^8 A^8   \\ 
&\quad{}+c_{48}^6 W^6 + c_{48}^4 W^4\ ,\\
W^6 \star W^6 &\sim c_{66}^{10} W^{10} + a_{66}^{10,1} A^{10,1} +  a_{66}^{10,2} A^{10,2} + c_{66}^8 W^8 + a_{66}^8 A^8 +c_{66}^6 W^6  \\
&\quad{}+ c_{66}^4 W^4  + n_{6} I\ ,
\end{split}
\end{equation}
where we have only written out the contributions of the Virasoro primaries to the singular part of the OPEs. 
(As mentioned before, the conformal symmetry fixes the contributions of their Virasoro descendants uniquely.) 
Furthermore, $A^8$, $A^{10,1}$, $A^{10,2}$, $A^{11}$
are the  composite primary fields at level 8, 10, 11, respectively, as predicted by 
(\ref{eq:Palg}). 
They are of the form
\begin{align} \label{eq:composites}
A^8 &= \left(W^4\right)^2+\cdots \ ,&
A^{10,1} &= {W^4}'' W^4 - \frac{9 (48+c)}{8(64+c)} {W^4}'{W^4}'+ \cdots\ ,\notag\\ 
A^{10,2} &= W^4 W^6 +\cdots \ ,&
A^{11} &= W^4 {W^6}' - \frac{3}{2} {W^4}'W^6+\cdots \ .
\end{align}
Here the ellipses denote Virasoro descendants that have to be added in order to make these fields
primary.


\subsubsection{Structure constants}
\label{eq:associativity}


Next we want to determine the structure constants appearing in (\ref{eq:wope}) by requiring
the associativity of the multiple OPEs $W^{s_1}\star W^{s_2}\star W^{s_3}$. Note that in this calculation,
we need to work with the full OPEs, rather than just their singular parts. The full OPE is in principle
uniquely determined  by its singular part, but the actual calculation is somewhat tedious. To do 
these computations efficiently we have therefore used the Mathematica packages 
\texttt{OPEdefs} and \texttt{OPEconf} of Thielemans that are described in some detail in 
{\cite{Thielemans:1991,Thielemans:1994}.\footnote{We thank K.~Thielemans for providing us 
with the packages.}  

More explicitly, we start by defining the OPE $W^4\star W^4$ by the first line of (\ref{eq:wope}), which
does not contain any composite fields. We can then use this ansatz to compute the composite field $A^8$,
and thus make an ansatz for the OPE $W^4\star W^6$. 
At this step, we can already  check the associativity of $W^4\star W^4 \star W^4$, using the 
built-in-function \texttt{OPEJacobi}. 

The next step consists of computing the composite fields made from $W^4$ and $W^6$, i.\,e.\ the 
remaining composite fields in \eqref{eq:composites}. Then we can make an ansatz for the remaining
OPEs in \eqref{eq:wope}, and check the associativity of $W^4\star W^4 \star W^6$.

It should now be clear how we continue: in each step we first compute all the composite primary fields
made of products of fundamental fields whose OPE we have already determined. This then
allows us to make an ansatz for the next `level' of OPEs. Then we can check the associativity of 
those triple products where all intermediate OPEs are known. Proceeding in this manner, we have 
computed the constraints arising from the associativity of the OPEs 
$W^{s_1}\star W^{s_2}\star W^{s_3}$ up to the total level $s_1+s_2+s_3\leq 16$. 
The resulting relations are (for the sake of brevity we only give the explicit expressions 
up to total spin $s_1+s_2+s_3\leq 14$ that can be calculated from the OPEs given 
explicitly in \eqref{eq:wope})

\begin{align}\label{eq:jacobi}
n_4 & = \tfrac{c (c-1) (c+24) (5 c+22) }{12 (2 c-1)(7 c+68)^2} \left(c_{46}^6\right)^2 - \tfrac{7 c (c-1) (5 c+22)}{72 (2 c-1)(7 c+68)} c_{46}^6 c_{44}^4 +
\tfrac{c (5 c+22)}{72 (c+24)} \left( c_{44}^4 \right)^2 \ , \notag\\ \notag
c_{44}^6 c_{46}^4 & = - \tfrac{8 (c-1) (c+24) (5 c+22) \left(c^2-172 c+196\right)}{(2 c-1)^2 (7 c+68)^3} \left( c_{46}^6 \right)^2 
+\tfrac{28 (c-1) (5 c+22) \left(c^2-172 c+196\right)}{3 (2 c-1)^2 (7 c+68)^2} c_{44}^4 c_{46}^6 \\ \notag
&\quad{}+\tfrac{4 (c-1) (5 c+22) (11 c+656)}{9 (c+24) (2 c-1) (7 c+68)} \left( c_{44}^4 \right)^2 \ , \\ \notag
c_{46}^8 a_{48}^{11} & =  \left( \tfrac{888}{65 c+2580}-\tfrac{40}{7 c+68} \right) c_{46}^6
- \tfrac{2 (13 c+918)}{65 c+2580} c_{44}^6 a_{46}^8 
+ \tfrac{224}{15 (c+24)} c_{44}^4 \ , \\ \notag
c_{48}^8 & = \tfrac{192-31 c}{26 c+1032}  c_{44}^6 a_{46}^{8}
+\tfrac{8 (c (33 c+1087)+11760)}{(7 c+68) (13 c+516)} c_{46}^6
-2 c_{44}^4 \ , \displaybreak[3]\\ \notag
c_{44}^6 c_{46}^8 a_{48}^8 & = 
\tfrac{192-31 c}{26 c+1032} \left( c_{44}^6 \right)^2 \left( a_{46}^{8} \right)^2 
-\tfrac{4 \left(165 c^3+10763 c^2+140036 c+38568\right)}{3 (c+24) (c+31) (55 c-6)} c_{44}^4 c_{44}^6 a_{46}^{8} \\ \notag
&\quad{}+ \tfrac{8 (c (33 c+1087)+11760)}{(7 c+68) (13 c+516)} c_{44}^6 a_{46}^{8} c_{46}^6
-\tfrac{896 (3 c+46) (5 c+3) ((c-172) c+196)}{(c+31) (2 c-1) (7 c+68)^2 (55 c-6)} \left( c_{46}^6 \right)^2 \\ \notag
&\quad{}+ \tfrac{3136 (3 c+46) (5 c+3) ((c-172) c+196)}{3 (c+24) (c+31) (2 c-1) (7 c+68) (55 c-6)} c_{44}^4 c_{46}^6 
+ \tfrac{448 (3 c+46) (5 c+3) (11 c+656)}{9 (c+24)^2 (c+31) (55 c-6)} (c_{44}^4)^2 \ ,\displaybreak[3] \\ \notag
c_{46}^8 c_{48}^6 & = 
-\tfrac{35 (c+50) (2 c-1) (7 c+68)}{3 (c+24) (c+31) (55 c-6)} c_{44}^4 c_{44}^6 a_{46}^{8}
+  \tfrac{8 \left(25 c^3+615 c^2-88272 c+102332\right)}{3 (c+24) (c+31) (55 c-6)} c_{44}^4 c_{46}^6 \\ \notag
&\quad{}+  \tfrac{16 \left(425 c^4+15145 c^3+233766 c^2+6507708 c-7565544\right)}{(c+31) (7 c+68) (13 c+516)
   (55 c-6)} \left( c_{46}^6 \right)^2 
+ \tfrac{604-4 c}{13 c+516} c_{44}^6 c_{46}^6 a_{46}^8 \\ \notag
&\quad{}+ \tfrac{7840 (c+50) (2 c-1) (7 c+68)}{9 (c+24)^2 (c+31) (55 c-6)} \left( c_{44}^4 \right)^2 \ , \\ \notag 
c_{44}^6 c_{46}^8 c_{48}^4 & = -\tfrac{(c-1) (c+24) (5 c+22) \left(65 c^4+8637 c^3+364470 c^2+2897944 c+36384\right)}{2 (2 c-1)
   (3 c+46) (5 c+3) (7 c+68)^2 (13 c+516)} c_{44}^6 a_{46}^8 \left( c_{46}^6 \right)^2 \\ \notag
&\quad{}+ \tfrac{7 (c-1) (5 c+22) \left(65 c^4+8637 c^3+364470 c^2+2897944 c+36384\right)}{12 (2 c-1) (3
   c+46) (5 c+3) (7 c+68) (13 c+516)} c_{44}^6 a_{46}^8 c_{44}^4 c_{46}^6 \\ \notag
&\quad{}-\tfrac{5 (c-1) (c+50) (5 c+22) \left(715 c^4+90933 c^3+2851076 c^2+21154896 c+6967008\right)}{12
   (c+24) (c+31) (3 c+46) (5 c+3) (13 c+516) (55 c-6)} \left( c_{44}^4 \right)^2 c_{44}^6 a_{46}^8 \\ \notag
&\quad{}-\tfrac{32 (c-151) (c-1) (c+24) (5 c+22) \left(c^2-172 c+196\right)}{(2 c-1)^2 (7 c+68)^3 (13
   c+516)} \left( c_{46}^6 \right)^3 \\ \notag
&\quad{}-\tfrac{56 (c-1) (5 c+22) \left(c^2-172 c+196\right) \left(20 c^3+24807 c^2+765640
   c-185172\right)}{3 (c+31) (2 c-1)^2 (7 c+68)^2 (13 c+516) (55 c-6)} c_{44}^4 \left( c_{46}^6 \right)^2 \\ \notag
&\quad{}+ \tfrac{4 (c-1) (5 c+22) \left(5605 c^4-408494 c^3-70820464 c^2-1703657536 c+1312613664\right)}{9
   (c+24) (c+31) (2 c-1) (7 c+68) (13 c+516) (55 c-6)} \left( c_{44}^4 \right)^2 c_{46}^6 \\ \notag
&\quad{}+ \tfrac{140 (c-1) (c+50) (5 c+22) (11 c+656)}{27 (c+24)^2 (c+31) (55 c-6)} \left( c_{44}^4 \right)^3
\ ,\\ \notag
c_{44}^6 c_{66}^{10} & = \tfrac{3}{4} {c_{46}^8} {c_{48}^{10}} \ , \\ \notag
\left( c_{44}^6 \right)^2 a_{66}^{10,1} &= 
\tfrac{3}{4} c_{44}^6 c_{46}^8 a_{48}^{10,1} +
\tfrac{5 (c+64) (c+76) (5 c+22) (11 c+232)}{3 (c+24) (c+31) (17 c+944) (55 c-6)} c_{44}^6 a_{46}^{8} c_{44}^4 \\ \notag
&\quad{}+ \tfrac{1120 (c+64) (11 c+656) (47 c-614)}{9 (c+24)^2 (c+31) (17 c+944) (55 c-6)} \left( c_{44}^4 \right)^2 
 \\ \notag
&\quad{}+\tfrac{7840 (c+64) (47 c-614) \left(c^2-172 c+196\right)}{3 (c+24) (c+31) (2 c-1) (7 c+68) (17 c+944) (55 c-6)} c_{46}^6 c_{44}^4 \\ \notag
&\quad{}-\tfrac{2240 (c+64) (47 c-614) \left(c^2-172 c+196\right)}{(c+31) (2 c-1) (7 c+68)^2 (17 c+944) (55 c-6)} \left( c_{46}^6 \right)^2 \ , \\ \notag
c_{44}^6 a_{66}^{10,2} & = \tfrac{6 (c+64) (13 c+248)}{(13 c+516) (17 c+944)}
c_{44}^6 a_{46}^{8} + \tfrac{192 (c+64) (81 c+1274)}{(7 c+68) (13 c+516) (17 c+944)}
c_{46}^6  -  \tfrac{224 (c+64)}{(c+24) (17 c+944)} c_{44}^4 \\ \notag
&\quad{}+  \tfrac{3}{4}  c_{46}^8 a_{48}^{10,2}  \ , \\ \notag
c_{44}^6 c_{66}^8 & = \tfrac{4 (4 c+61)}{7 c+68} c_{46}^8 c_{46}^6 -\tfrac{(11 c+656)}{6 (c+24)} c_{46}^8 c_{44}^4 \ , \\ \notag
\left( c_{44}^6 \right)^2 a_{66}^8 & = 
-\tfrac{11 c+656}{6 (c+24)} c_{44}^4 c_{44}^6 a_{46}^{8}
+ \tfrac{784 ((c-172) c+196)}{3 (c+24) (2 c-1) (7 c+68)} c_{44}^4 c_{46}^6
-\tfrac{224 ((c-172) c+196)}{(2 c-1) (7 c+68)^2} \left( c_{46}^6 \right)^2 \\ \notag
&\quad{}+ \tfrac{4 (4 c+61)}{7 c+68} c_{46}^6 c_{44}^6 a_{46}^8 
+ \tfrac{112 (11 c+656)}{9 (c+24)^2} \left( c_{44}^4 \right)^2 \ , \displaybreak[3]\\ \notag
c_{44}^6 c_{66}^6 & = 
\tfrac{20 \left(92 c^5+2389 c^4+39632 c^3+4060 c^2-212032 c+193984\right)}{(2 c-1)^2 (7 c+68)^3} 
\left( c_{46}^6 \right)^2 \\ \notag
&\quad{}+ \tfrac{10 \left(28 c^5-5425 c^4-525974 c^3+387728 c^2+3726976 c-3870208\right)}{9 (c+24) (2 c-1)^2 (7 c+68)^2} 
c_{46}^6 c_{44}^4  \\ \notag
&\quad{}-\tfrac{20 \left(13 c^4-1637 c^3-113622 c^2+32168 c+859328\right)}{27 (c+24)^2 (2 c-1) (7 c+68)} 
\left( c_{44}^4 \right)^2 \ , \displaybreak[4]\\ \notag
\left( {c_{44}^6} \right)^2 c_{66}^4 & = -\tfrac{8 (c-1) (c+24) (5 c+22) ((c-172) c+196)}{(1-2 c)^2 (7 c+68)^3} \left( c_{46}^6 \right)^3
 \\ \notag
&\quad{}+ \tfrac{28 (c-1) (5 c+22) ((c-172) c+196)}{3 (1-2 c)^2 (7 c+68)^2} c_{44}^4 \left( c_{46}^6 \right)^2
+ \tfrac{4 (c-1) (5 c+22) (11 c+656)}{9 (c+24) (2 c-1) (7 c+68)} \left( c_{44}^4 \right)^2 c_{46}^6  \ , \displaybreak[3]\\ \notag
\left( c_{44}^6 \right)^2 n_{6} & = -\tfrac{2 (c-1)^2 c (c+24)^2 (5 c+22)^2 ((c-172) c+196)}{3 (2 c-1)^3 (7 c+68)^5} \left(c_{46}^6\right)^4 + \tfrac{(c-1) c (5 c+22)^2 (11 c+656)}{162 (c+24)^2 (2 c-1) (7 c+68)} \left( c_{44}^4 \right)^4 \\ \notag
&\quad{}+ \tfrac{14 (c-1)^2 c (c+24) (5 c+22)^2 ((c-172) c+196)}{9 (2 c-1)^3 (7 c+68)^4} c_{44}^4 \left( c_{46}^6 \right)^3 \\ \notag
&\quad{}-\tfrac{(c-1) c (5 c+22)^2 (c (c (17 c-13105)+25330)-12092)}{54 (2 c-1)^3 (7 c+68)^3} \left(c_{44}^4\right)^2 \left(c_{46}^6\right)^2 \\ 
&\quad{}-\tfrac{7 (c-1) c (5 c+22)^2 (c (8 c+1161)-1244)}{162 (1-2 c)^2 (c+24) (7 c+68)^2} \left( c_{44}^4 \right)^3 c_{46}^6\ .
\end{align}
Let us comment on the implications of these results. Of the 
$23$ structure constants that appear in \eqref{eq:wope}, $8$ remain undetermined by the
above relations; for example, a convenient choice for the free structure constants is 
$n_4$, $n_6$, $c_{46}^8$, $c_{48}^{10}$, as well as $a_{46}^{8}$, $a_{48}^{10,1}$, $a_{48}^{10,2}$ and
$c_{44}^4$. The first $4$ of these just account for the freedom to normalise the
fields $W^4, W^6, W^8$ and $W^{10}$, respectively. The appearance of 
$a_{46}^{8}$, $a_{48}^{10,1}$, and $a_{48}^{10,2}$ also has a simple interpretation: it 
reflects the freedom to redefine $W^8$ and $W^{10}$ by adding to them composite fields of the same spin. 
For  example, we can set $\hat{a}_{46}^{8}=0$ by redefining 
$W^8\mapsto \hat{W}^8\equiv  W^8 + a_{46}^{8}/ c_{46}^{8} A^8$, and similarly in the other two cases. 
Note that this freedom implies that the relations (\ref{eq:jacobi}) must satisfy interesting consistency
conditions. For example, if we redefine $\hat{W}^8$ in this manner, the structure constant $a_{48}^{11}$
in the OPE $W^4\star \hat{W}^8$  becomes 
$\hat{a}_{48}^{11} = a_{48}^{11}+\tfrac{a_{46}^{8}}{c_{46}^{8}}\, \tfrac{2(13c+918)}{65c+2580} c_{44}^6$,
which then satisfies indeed the third equation of (\ref{eq:jacobi}) with $\hat{a}_{46}^{8}=0$. 

Thus, at least up to the level to which we have analysed the Jacobi identities and up to 
field redefinitions, all structure constants
are completely fixed in terms of $c$ and the single fundamental structure constant $c_{44}^4$. Note that
for a given choice of $n_4$, the sign of  $c_{44}^{4}$ is not determined since $n_4$ only fixes the
normalisation of $W^4$ up to a sign. 
It seems reasonable to believe that this structure will continue, i.\,e.\ that all remaining structure
constants are also uniquely fixed (up to field redefinitions) in terms of the central charge $c$ and 
\begin{equation}
\gamma =  \left( c_{44}^4 \right)^2  \ .
\end{equation}
The situation is then analogous to what was found for ${\cal W}_{\infty}[\mu]$ in \cite{Gaberdiel:2012ku},
and for $s{\cal W}_{\infty}[\mu]$ in \cite{Candu:2012tr}: the resulting algebra depends on one 
free parameter (in addition to the central charge $c$), and whenever we want to emphasise this
dependence, we shall denote it by $\We(\gamma)$. 

In a next step we want to relate $\We(\gamma)$ to the Drinfel'd-Sokolov reduction of 
$\hse$. Before doing so,
we can however already perform some simple consistency checks on the above analysis.


\subsection{Truncations}
\label{sec:trunc}


Since our ansatz is completely general, it should also reproduce the various finite even ${\cal W}$ 
algebras that have been constructed  in the literature before
\cite{KauschWatts:1990,Blumenhagen:1990jv}. More specifically, we can study
for which values of $\gamma$, $\mathcal{W}_{\infty}^e$  develops an ideal such that 
the resulting quotient algebra becomes a finite ${\cal W}$ algebra.

\subsubsection{The algebra \texorpdfstring{${\cal W}(2,4)$}{W(2,4)}}


The simplest case is the so-called ${\cal W}(2,4)$ algebra, which is generated by a single 
Virasoro primary field $W^4$ in addition to the stress energy tensor. Thus we need to find
the value of $\gamma$ for which $W^6$, $W^8$, etc.\ lie in an ideal. Imposing 
$c_{46}^4$, $c_{66}^4$, $c_{48}^4$, and $n_6$ to vanish we obtain 
\begin{equation}
\label{eqn:gamma24}
 \gamma = \frac{216(c+24)(c^2-172c+196)n_4}{c(2c-1)(7c+68)(5c+22)}\ .
\end{equation}
The resulting quotient algebra is then in agreement with e.\,g.\ \cite{KauschWatts:1990}. Note 
that $A^8$ does not lie in the ideal since the OPE of 
$W^4$ with $A^8$ contains terms proportional to $W^4$ which are non-vanishing for 
generic $c$. Thus we also need to require that $a_{46}^{8}=0$ (but $c_{46}^{8}$ need
not be zero), which is automatically true by the above conditions. We have also analysed the consistency of the resulting algebra directly, i.\,e.\
repeating essentially the same calculation as in \cite{KauschWatts:1990}.


\subsubsection{The algebras \texorpdfstring{${\cal W}(2,4,6)$}{W(2,4,6)}}
\label{sec:w246}


The next simplest case is the so-called ${\cal W}(2,4,6)$ algebra, which should appear from
${\cal W}^{e}_{\infty}$ upon dividing out the ideal generated by $W^8$, $W^{10}$, etc. This
requires that we set $c_{48}^4, c_{48}^6$ and $n_8$ to zero.
Furthermore, since the composite fields $A^{8}$, $A^{10,1}$, $A^{10,2}$ and $A^{11}$ have 
a non-trivial image in the quotient (for generic $c$), we should expect that also 
$a_{48}^8$, $a_{48}^{10,1}$, $a_{48}^{10,2}$ and $a_{48}^{11}$ vanish. 
Solving eqs.~\eqref{eq:jacobi} together with these constraints then yields the two values for $\gamma$
\begin{align}\label{eq:gamma246}
 \gamma &= 2n_4\Bigl[ (18025c^6 + 1356090c^5+16727763c^4-537533674c^3 \notag\\
 &\quad\qquad\qquad{}-5470228116c^2+8831442312c -300564000)\notag\\
 &\quad\qquad {}\pm (c-1)(5c+22)^2(11c+444)(13c+918)\sqrt{c^2-534c+729}\Bigr]\notag\\
&\quad\times\bigl[c(2c-1)(3c+46)(4c+143)(5c+3)(5c+22)(5c+44)\bigr]^{-1}\ .
\end{align}
Up to a factor of $\frac{1}{2}$, this agrees with two of the four solutions found in
\cite{KauschWatts:1990}; 
incidentally, they are the ones which were claimed to be inconsistent in \cite{Kliem}.
We have 
again also analysed the consistency of the resulting algebra directly, i.\,e.\  by working with
an ansatz involving only $W^4$ and $W^6$.

As a matter of fact, there are two additional solutions that appear if we 
enlarge the ideal by also taking $A^{11}$ to be part of it. 
Then we do not need to impose that 
$a_{48}^{11}=0$, and the resulting algebras agree with the other two solutions\footnote{Incidentally, there is a 
typo in \cite{KauschWatts:1990}: the structure constant $a_{46}^8$ in the ${\cal W}(2,4,6)$
algebra satisfying \eqref{eq:gamma_set1} should be given by 
$(a_{46}^8)^2=\frac{256(2c-1)(5c+3)^2(3c+46)^2(7c+68)n_6}{(31c-192)^2(c+11)(14c+11)(5c+22)(c+24)n_4}$\ .}
of \cite{KauschWatts:1990}, i.\,e.\ they are  characterised by
\begin{equation}
\label{eq:gamma_set2}
 \gamma = -\frac{4 (5 c^2+309c-14)^2 n_4}{c(c-26) (5 c+3) (5 c+22)}
\end{equation}
and
\begin{equation}
\label{eq:gamma_set1}
 \gamma = \frac{216 (10 c^2+47c-82)^2 n_4}{c(4 c+21) (5 c+22) (10 c-7)}\ ,
\end{equation}
respectively. In the case of \eqref{eq:gamma_set2}, the OPEs of $W^4$ and $W^6$ with $A^{11}$ show
that no additional field of dimension smaller than 11 needs to be included in the ideal.
However, in the case of the algebra described by \eqref{eq:gamma_set1}, the ideal also contains a
certain linear combination of $A^{10,1}$ and $A^{10,2}$.

\subsection{Identifying the wedge algebra}
\label{sec:wedge}

We expect from the analysis of \cite{Gaberdiel:2011nt} that $\We(\gamma)$ should 
arise as the Drinfel'd-Sokolov reduction of the even higher spin algebra. However,
as was also explained in  \cite{Gaberdiel:2011nt}, it is not clear which higher spin
algebra is relevant in this context, and two possibilities, ${\rm hs}^{e}[\mu]$ and
${\rm hso}[\mu]$, were proposed. In order to decide which of the two algebras is relevant,
it is sufficient to determine the wedge algebra of $\We(\gamma)$, i.\,e.\ the algebra that
is obtained by restricting the modes to the wedge $\lvert m \rvert < s$, and
taking $c\rightarrow \infty$. (The reason for this is that restricting to the wedge algebra is in a sense
the inverse of performing the Drinfel'd-Sokolov reduction, see \cite{Bowcock:1991zk,Gaberdiel:2011wb} for a discussion 
of this point.)
As it turns out, the 
wedge algebra commutators of $\We(\gamma)$, as obtained from \eqref{eq:wope} together with (\ref{eq:jacobi}),
agree with the $\hse$ commutators (\ref{hsecom}) of appendix~\ref{sec:hse} provided we identify
$c_{44}^{4}=\sqrt{\gamma}$ with $\mu$ as
\begin{equation}\label{eq:c444}
c_{44}^4 = \frac{12}{\sqrt{5}}(\mu^2-19) + \mathcal{O}(c^{-1}) \ .
\end{equation}
Furthermore, we normalise our fields as 
\begin{align}\label{eq:n4}
n_4 &= c (\mu^2-9)(\mu^2-4)\ ,& n_6 &= c (\mu^2-25)(\mu^2-16)(\mu^2-9)(\mu^2-4)\ ,\\ \notag
c_{46}^8&=-8\sqrt{\tfrac{210}{143}}\ ,& c_{48}^{10}&=-20 \sqrt{\tfrac{6}{17}}\ ,
\end{align}
and take the field redefinition parameters to be 
$a_{46}^{8}=a_{48}^{10,1}=a_{48}^{10,2}=0$. Thus we conclude that the $\We(\gamma)$ algebra can be interpreted as
the quantum Drinfel'd-Sokolov reduction of $\hse$, where $\mu$ and $\gamma$ are related as in (\ref{eq:c444}); this
will be further elaborated on in section~\ref{sec:ds}.

We should mention in passing that $\hse$ and ${\rm hso}[\mu']$ are not isomorphic (even allowing 
for some general relation between $\mu$ and $\mu'$), since they possess different finite-dimensional
quotient algebras, see~\cite{Gaberdiel:2011nt}. Thus the above analysis also proves that the wedge algebra
of $\We$ is {\em not} isomorphic to ${\rm hso}[\mu]$ for any $\mu$, and hence that $\We$ is {\em not} the
quantum Drinfel'd-Sokolov reduction of ${\rm hso}[\mu]$ for any $\mu$.

\subsection{Minimal representation}
\label{sec:min}

Our next aim is to determine the exact $c$ dependence of (\ref{eq:c444}). This can be done
using the same trick as in \cite{Gaberdiel:2012ku} and \cite{Candu:2012tr}, following the original analysis of 
 \cite{Hornfeck:1993kp}. The main ingredient
in this analysis is a detailed understanding of the structure of the `minimal representations' of 
$\We$. Recall that the duality of \cite{Gaberdiel:2011nt} suggests that $\We$ possesses two minimal
representations  whose character is of the form 
\begin{equation}\label{minchar}
\chi_{\text{min}}(q) = \frac{q^h}{1-q} \prod_{s \in 2\mathbb{N}} \prod_{n=s}^{\infty} \frac{1}{1-q^n} \ ,
\end{equation}
and for which $h$ is finite in the 't~Hooft limit. It follows from this character formula that the 
corresponding representation has (infinitely)
many low-lying null-vectors; this will allow us to calculate $h$ as a function of $c$ and $\gamma$. 
\smallskip

Let us denote the primary field of the minimal representation by $P^0$. First, we need to 
make the most general ansatz for the OPEs $W^s\star P^0$. In order to do so we have
to enumerate the number of Virasoro primary states in the minimal $\We$  representation. 
Decomposing (\ref{minchar}) in terms of irreducible Virasoro characters as 
\be\label{minde}
\chi_{\rm min}(q) = \sum_{n=0}^{\infty} d_{\rm min}(n) \, \chi_{h+n}(q) \ , 
\ee
where $\chi_h(q)$ was defined in (\ref{chih}), $d_{\rm min}(n)$ equals then the multiplicity of 
the Virasoro primaries of conformal dimension $h+n$. It follows from (\ref{minde}) 
that the corresponding generating function is 
\begin{equation}\label{minexp}
\sum_{n=0}^\infty d_{\text{min}}(n) \, q^{n} = \prod_{s=2}^{\infty} \prod_{n=2s}^{\infty} \frac{1}{1-q^n} 
= 1+ q^4 + q^5 + 2q^6 
+\cdots\ .
\end{equation}
Then the most general ansatz 
for the OPEs $W^{4} \star P^0$ and $W^{6} \star P^0$ is
\begin{align}\label{eq:opeWP}
W^4 \star P^0 &\sim w^4 P^0\ ,&
W^6 \star P^0 &\sim w^6 P^0 + a^4 P^4 + a^5 P^5 \ ,
\end{align}
where $P^4$ and $P^5$ are the primary fields of conformal dimension $h+4$ and $h+5$, respectively.
Note that these fields are unique, as follows from (\ref{minexp}); explicitly, they are of the form
\begin{align}
P^4 &= W^4 P^0+\cdots \ ,& P^5 = \frac{h}{4+h} {W^4}' P^0 - \frac{4}{4+h} W^4 {P^0}'+\cdots\ ,
\end{align}
where the ellipses stand for Virasoro descendants that are required to make these fields primary. As in 
\cite{Candu:2012tr}, the condition
that $P^0$ defines a representation of $\We$ is now equivalent to the constraint that all OPEs
$W^{s_1} \star W^{s_2} \star P^0$ are associative. While we cannot test all of these conditions, 
imposing the associativity of $W^4\star W^4\star P^0$ implies already \pagebreak[3]
\begin{align}
w^4 &=\frac{12 h \left(c^2 (9-2 (h-1) h)+3 c (h ((49-12 h) h-40)+2)-2 h (h (12 h+5)-14)\right)}{( c (h-2) (2 h-3)+h (4 h-5))}\notag\\ 
&\quad{}\times \frac{n_4}{c(5 c+22)c_{44}^4}\ ,\notag\\
w^6 &= \frac{8 (c-1) (5 c+22) h \left(c (h+2)+15 h^2-26 h+8\right) (c (2 h+3)+4 h (12 h-7))n_4}{3 c(c+24) (2 c-1)
 (7c+68) (c (h-2) (2 h-3)+h (4 h-5))c_{44}^6}\ , \notag \\
a^4 
& = \frac{16 (5 c+22) (4 h-9) \left(c (h+2)+15 h^2-26 h+8\right)}{(c+24) \left((c-7) h+c+3 h^2+2\right) (2 c h+c+2 h (8 h-5)) c_{44}^6} w^4\ ,  \notag \\
a^5 &  = 
\frac{20 (5 c+22) (h-4) (h-1) (c (2 h+3)+4 h (12 h-7))}{(c+24) h \left((c-7) h+c+3 h^2+2\right) 
(2 c h+c+2 h (8 h-5)) c_{44}^6} w^4 \ , \label{eq:opeP} 
 \end{align}
up to a sign ambiguity of the self-coupling $c_{44}^4=\pm \sqrt{\gamma}$.
Furthermore, the conformal dimension $h$ is determined by the equation 

\begin{align}\label{eqn:gamma0}
\gamma&=\frac{[c^2 (-9 - 2 h + 2 h^2) + 
 3 c (-2 + 40 h - 49 h^2 + 12 h^3) +2 h (-14 + 5 h + 12 h^2) ]^2}
{ [c(1+h)+2  - 7 h +  3 h^2] [(1+2h)c - 10 h  + 
   16 h^2] [(6 -7h+2h^2)c - 5 h + 4 h^2]} \notag\\
& \quad\times\frac{144 n_4}{c (5c+22)}\ .
\end{align}
Given $\gamma$ and $c$, this is a sextic equation for $h$. We also note that our  result 
is consistent with the one obtained in \cite{Hornfeck:1993kp}. Moreover, we have checked 
that we arrive at the same result using commutators instead of OPEs; this calculation, which is 
analogous to the one performed  in \cite{Gaberdiel:2012ku} for the algebra $\mathcal{W}_\infty[\mu]$, is 
presented in appendix~\ref{sec:commutators}.

We should stress that the above constraints are necessary conditions for the minimal representation
to exist, but do not prove that they are actually compatible with the full $\We$ structure. Furthermore, 
since we have only used the low-lying OPEs, our analysis actually holds for  \emph{any} algebra of type 
$\mathcal{W}(2,4,\dots)$ with no simple field of spin $5$, and for any representation whose character
coincides with  (\ref{minchar}) up to $q^5$, see also \cite{Hornfeck:1993kp}.


\section{Drinfel'd-Sokolov reductions} \label{sec:ds}


As we have seen in section \ref{sec:wedge}, the wedge algebra of $\We(\gamma)$ is $\hse$, where
$\gamma=(c_{44}^{4})^2$ is identified with a certain function of $\mu$, see eq.~(\ref{eq:c444}).
Thus we should expect that the quantum $\We[\mu]$ algebras (where we now label $\We$ in terms
of $\mu$ rather than $\gamma$) can be thought of as being the 
Drinfel'd-Sokolov reduction of $\hse$. Actually, as we shall shortly see, the situation is a little bit more 
complicated. The subtlety we are about to encounter is 
related to the fact that $\hse$ is in some sense a non-simply-laced 
algebra.\footnote{Note, however, the situation is also complicated by the fact that 
$\hse$ is infinite dimensional, and the construction of \cite{Bouwknegt:1992wg}  only applies 
to finite-dimensional Lie algebras. On the other hand, given that things worked nicely \cite{Gaberdiel:2012ku}
for the infinite-dimensional algebra ${\rm hs}[\mu]$, we suspect that 
the infinite-dimensionality of $\hse$ is not the origin of the subtlety.}

Since Drinfel'd-Sokolov reductions of infinite-dimensional Lie algebras are complicated, we shall
first (as in \cite{Gaberdiel:2012ku}) consider the special cases when $\mu$ is a positive integer. 
Then $\hse$ can be reduced to finite-dimensional Lie algebras; 
indeed, as was already explained in \cite{Gaberdiel:2011nt}, we have 
\begin{equation}\label{eq:redhse}
\mathrm{hs}^e[N]/\chi_N = \begin{cases}
\mathfrak{so}(N) & \text{for $N$ odd}\\
\mathfrak{sp}(N) & \text{for $N$ even,}
\end{cases}
\end{equation}
where $\chi_N$ is the ideal of $\hse$ that appears for $\mu=N\in\mathbb{N}$. Note that in both
cases, the resulting algebra is non-simply-laced, suggesting that $\hse$ should be thought of as
being non-simply-laced itself. 

As in \cite{Gaberdiel:2012ku} we should now expect that the quantum Drinfel'd-Sokolov reduction of $\hse$ agrees, 
for $\mu=N$, with the quantum Drinfel'd-Sokolov reduction of \linebreak
$B_n=\mathfrak{so}(2n+1)$ or $C_n=\mathfrak{sp}(2n)$, respectively. 
The representation theory of these ${\cal WB}_n$ and ${\cal WC}_n$ algebras is well known,
and thus, at least for these integer values of $\mu$, we can compare the conformal dimension
of the corresponding minimal representations with what was determined above, see eq.~(\ref{eqn:gamma0}). 
This will allow us to deduce an exact relation between $\gamma$ 
and $\mu=N$ (for all values of the central charge). Analytically continuing the resulting expression
to non-integer $\mu$ should then lead to the precise relation between $\gamma$ and $\mu$,
for all values of $\mu$.

\subsection{The \texorpdfstring{$B_n$}{Bn} series approach}
\label{sec:bds}

According to~\cite{Bouwknegt:1992wg},  the Drinfel'd-Sokolov reduction of $\mathfrak{so}(2n+1)$, 
which we shall denote by $\mathcal{WB}_n$, is an algebra of type 
$\mathcal{W}(2,4,\dots, 2n)$ with central charge
\begin{equation}\label{eq:cds}
c _B= n - 12 \vert \alpha_+ \rho_B + \alpha_- \rho_B^{\vee} \vert^2  
\end{equation}
and spectrum
\begin{equation}\label{eq:hds}
h_{\Lambda}= \tfrac{1}{2}(\Lambda, \Lambda + 2\alpha_+ \rho_B + 2\alpha_-\rho_B^\vee)\ ,
\qquad \Lambda\in \alpha_+ P_+ + \alpha_- P^\vee_+\ .
\end{equation}
Here $\rho_B$ and $\rho_B^\vee$ are the $\mathfrak{so}(2n+1)$ Weyl vector and covector, respectively, 
and $P_+$ and $P^\vee_+$  are the lattices of $\mathfrak{so}(2n+1)$
dominant weights and coweights, respectively. We work with the convention that the long roots have length 
squared equal to $2$,  and $\alpha_\pm$ are defined by 
\begin{equation}\label{eq:aminus}
\alpha_- = -\sqrt{k_B+2n-1}\ ,\qquad \alpha_+ = \frac{1}{\sqrt{k_B+2n-1}} 
\ ,
\end{equation}
so that $\alpha_+\alpha_-=-1$. Furthermore $k_B$ is the level that appears in the Drinfel'd-Sokolov
reduction. Note that the dual Coxeter number of 
$\mathfrak{so}(2n+1)$ equals $g_B=2n-1$. Plugging in the expressions for $\alpha_\pm$ into
\eqref{eq:cds}, the central charge  of ${\cal WB}_n$ takes the form
\begin{eqnarray}\label{eq:redcB}
c_B(\mu,k_B) & = &  -\frac{n [k_B(2n+1)+4 n^2-2n] [2 k_B (n+1)+4 n^2-3]}{k_B+2 n-1}\\
& = &  \frac{(1-\mu) (k_B \mu +\mu^2  - 3 \mu + 2) [k_B(1+\mu)+\mu^2 - 2 \mu -2]}{2 (k_B +\mu-2)}\ ,\nonumber
\end{eqnarray}
where in the second line we have replaced $n=\tfrac{\mu-1}{2}$.
The minimal representations of ${\cal WB}_n$ arise for $\Lambda=\Lambda_+=\alpha_+ {\rm f}$,
and $\Lambda=\Lambda_-=\alpha_{-} {\rm f}^{\vee}$, where ${\rm f}$ is the highest weight of 
the fundamental $\mathfrak{so}(2n+1)$ representation, and ${\rm f}^{\vee}$ the corresponding coweight.
The conformal dimensions of these two representations are 
\begin{equation}\label{eq:hredmin}
h_+=h_{\Lambda_+} = -\frac{n (k_B+2 n-2)}{k_B+2 n-1}\ ,\qquad 
h_-=h_{\Lambda_-} = k_B \left(n+\tfrac{1}{2}\right)+n (2 n-1) \ .
\end{equation}
They are both solutions of eq.~\eqref{eqn:gamma0}, provided  $\gamma=\gamma_B(\mu,k_B)$ 
with $\gamma_B$ equal to
\begin{align}\label{eq:cgbk}
\gamma_B =& 
144\,  (240 - 420 k_B + 210 k_B^2 - 30 k_B^3 + 1188 \mu -  1520 k_B \mu + 773 k_B^2 \mu 
- 190 k_B^3\mu  \notag
\\ \notag &
+ 19 k_B^4 \mu - 2138 \mu^2 + 2237 k_B \mu^2 - 818 k_B^2 \mu^2 + 107 k_B^3 \mu^2 + 264 \mu^3 
+ 614 k_B \mu^3 \\ \notag &
-  703 k_B^2 \mu^3 + 220 k_B^3 \mu^3 - 20 k_B^4 \mu^3 + 1107 \mu^4 
- 1516 k_B \mu^4 + 615 k_B^2 \mu^4 -  75 k_B^3 \mu^4 
\\ \notag&
- 644 \mu^5 + 462 k_B \mu^5  - 51 k_B^2 \mu^5 - 12 k_B^3 \mu^5 + k_B^4 \mu^5
+ 67 \mu^6 + 43 k_B \mu^6 -  36 k_B^2 \mu^6  \\ \notag&
+ 4 k_B^3 \mu^6 + 39 \mu^7 - 36 k_B \mu^7 + 4 k_B \mu^8 + \mu^9 +  6 k_B^2 \mu^7 - 12 \mu^8 )^2 \, n_4\, 
/\, 
\Bigl[ c_B (\mu-3)
      \\ \notag&
\times   (3 k_B +  k_B \mu -6+ \mu^2)  (8 - 2 k_B - 5 \mu + k_B \mu + \mu^2)
(1 -  k_B - 4 \mu + k_B \mu + \mu^2)
 \\ \notag&
\times (1 - 3 \mu +  k_B \mu + \mu^2)  ( k_B+k_B\mu-4 -  2 \mu  + \mu^2) 
(2 k_B -2- \mu + k_B \mu + \mu^2) \\ & 
\times (108 - 54 k_B - 74 \mu + 25 k_B \mu - 5 k_B^2 \mu - 20 \mu^2 + 5 k_B \mu^2 + 55 \mu^3 
- 30 k_B \mu^3  
\notag
 \\ & \ \ 
\qquad + 5 k_B^2 \mu^3 - 30 \mu^4+ 10 k_B \mu^4    +  5 \mu^5) \Bigr]
\ ,
\end{align}
where we have again replaced $n=\tfrac{\mu-1}{2}$. For each $\mu$, we therefore obtain a 
family of $\We$ algebras that depend on $k_B$; these algebras will be denoted by $\WeB[\mu]$
(where we suppress the explicit $k_B$ dependence). Note  that, for fixed $\mu$, these algebras 
really depend on $k_B$, rather than just on $c_B$: for a fixed $c$ and $\mu$, there are always two 
solutions $k_B^{(i)}$, $i=1,2$, for $c=c(\mu,k_B^{(i)})$, see (\ref{eq:redcB}). However, in general
the corresponding $\gamma$ values do not agree, 
$\gamma_B(\mu,k_B^{(1)}) \neq \gamma_{B}(\mu,k_B^{(2)})$, and hence the two solutions for
$k_B$ do not lead to isomorphic $\We$ algebras. This is different than what happened for 
${\cal W}_{\infty}$ in \cite{Gaberdiel:2012ku}, and closely related to the fact that $\hse$ is non 
simply-laced, see below.

By construction, the algebras $\WeB[\mu]$ truncate, for $\mu=2n+1$, to $\mathcal{WB}_n$. (Note
that also $\mathcal{WB}_n$ depends actually on the level $k_B$, and not just on $c$.) 
We have also checked that, for $n=2$, $\gamma_B(2n+1,k_B)$ agrees with the 
$\gamma$ given in eq.~(\ref{eqn:gamma24}) at $c=c_B(2n+1,k_B)$. Similarly, for $n=3$, 
$\gamma_B(2n+1,k_B)$ agrees with the $\gamma$ of eq.~(\ref{eq:gamma246}) at $c=c_B(2n+1,k_B)$.
(For $n=3$, the two algebras corresponding to the two different solutions for $k_B$ correspond to the 
choice of the branch cut in the square root of eq.~(\ref{eq:gamma246}).)

\subsection{The \texorpdfstring{$C_n$}{Cn} series approach}

The analysis for the Drinfel'd-Sokolov reduction of $\mathfrak{sp}(2n)$, which we shall
denote by $\mathcal{WC}_n$, is essentially identical. Also ${\cal WC}_n$ is an
algebra of type $\W(2,4,\ldots, 2n)$, and its central charge equals 
\begin{equation}
c_C = n - 12 \vert \alpha_+ \rho_C + \alpha_- \rho_C^{\vee} \vert^2 \ ,
\end{equation}
where now $\rho_C$ and $\rho_C^{\vee}$ are the Weyl vector and covector of $\mathfrak{sp}(2n)$,
respectively. The spectrum is described by the analogue of eq.~(\ref{eq:hds}), where\footnote{In our 
conventions, the {\em short} roots of $C_n=\mathfrak{sp}(2n)$ have  length squared equal to $2$.}
\be
\alpha_- = -\sqrt{k_C + 2 n +2} \ , \qquad \alpha_+ = \frac{1}{\sqrt{k_C + 2n + 2}} \ .
\ee
Expressed in terms of $n$ and $k_C$, the central charge then takes the form
\begin{align}\label{cC}
c_C (\mu, k_C) &= -\frac{n [ k_C (2 n+ 1) +4 n^2 +4 n)] [ k_C (2 n-1)+4 n^2-3 ]}{k_C+2 n+2} \\ \nonumber
&= -\frac{\mu  [ (k_C+2) \mu +k_C +\mu ^2 ] [ k_C (\mu -1)+\mu ^2-3 ]}{2 (k_C +\mu +2)} \ , 
\end{align}
where we have, in the second line, replaced $n=\frac{\mu}{2}$. The conformal dimensions
of the minimal representations are now 
\be
h_{+}=h_{\Lambda_+} =  \frac{k_C (1-2n) -4 n^2+3}{2 k_C + 4 n + 4} \ , \qquad
h_{-}= h_{\Lambda_-} = n (k_C+2 n+1) \ ,
\ee
and they are both solutions of eq.~\eqref{eqn:gamma0} provided 
$\gamma=\gamma_C(\mu,k_C)$ equals
\begin{align}\label{eq:cgck}
\gamma_C =& \,144 (-224 - 520 k_C - 340 k_C^2 - 68 k_C^3 - 888 \mu -  1064 k_C \mu 
- 161 k_C^2 \mu + 114 k_C^3 \mu \notag \\ 
& + 19 k_C^4 \mu - 372 \mu^2+ 687 k_C \mu^2 + 946 k_C^2 \mu^2 +   227 k_C^3 \mu^2 
+ 730 \mu^3 + 1390 k_C \mu^3  \notag \\
&
+  377 k_C^2 \mu^3 - 100 k_C^3 \mu^3 - 20 k_C^4 \mu^3 + 553 \mu^4
+ 134 k_C \mu^4 - 315 k_C^2 \mu^4 -  85 k_C^3 \mu^4   \notag \\
&
 - 34 \mu^5 - 326 k_C \mu^5 - 129 k_C^2 \mu^5 + 4 k_C^3 \mu^5 + k_C^4 \mu^5 - 111 \mu^6 
- 83 k_C \mu^6 +  12 k_C^2 \mu^6  \notag \\ 
&
+ 4 k_C^3 \mu^6 - 19 \mu^7 + 12 k_C \mu^7 +  6 k_C^2 \mu^7 + 4 \mu^8
+ 4 k_C \mu^8 + \mu^9)^2 n_4\, / \Bigl[
c_C(\mu-2) \notag \\
& \times ( k_C \mu -5 - k_C +\mu^2)( k_C \mu- 3 k_C -5-2 \mu + \mu^2)
(k_C \mu +4 +  3 k_C + 4 \mu + \mu^2)
\notag\\ 
&  \times (k_C \mu -2 + k_C +  2 \mu + \mu^2) ( k_C \mu +4 + 2 k_C + 3 \mu + \mu^2) 
( k_C \mu -1+ \mu + \mu^2)  \notag 
\\ \notag& 
\times (-88 - 44 k_C - 44 \mu -  15 k_C \mu - 5 k_C^2 \mu - 30 \mu^2 - 25 k_C \mu^2 
-  15 \mu^3 + 10 k_C \mu^3 \notag \\
& \ \ 
\qquad + 5 k_C^2 \mu^3 + 10 \mu^4   +   10 k_C \mu^4+ 5 \mu^5 ) \Bigr] \ ,
\end{align}
where we have again replaced $n=\tfrac{\mu}{2}$. For each $\mu$, we therefore obtain a 
family of $\We$ algebras that depend on $k_C$; these algebras will be denoted by $\WeC[\mu]$
(where we suppress as before the explicit $k_C$ dependence). Again, these algebras actually depend
on $k_C$, rather than just $c_C$. By construction, $\WeC[\mu]$ has the property that it truncates 
to $\mathcal{WC}_n$  for $\mu=2n$. We have also checked that, for $n=2$, $\gamma_C(2n,k_C)$ agrees with the 
$\gamma$ of eq.~(\ref{eqn:gamma24}) at $c=c_C(2n,k_C)$. Similarly, for $n=3$, 
$\gamma_C(2n,k_C)$ agrees with the $\gamma$ of eq.~(\ref{eq:gamma246}) at $c=c_C(2n,k_C)$, where
again the two solutions for $k_C$ correspond to the two signs in front of the square root in
eq.~(\ref{eq:gamma246}).

\subsection{Langlands duality}

Naively, one would have expected that the two quantum algebras $\WeB[\mu]$ and $\WeC[\mu]$
should be equivalent, but this is not actually the case: if we fix $\mu$ and $c$, and determine 
$k_B^{(i)}$, $i=1,2$, and $k_C^{(j)}$, $j=1,2$,  by the requirement that
\be
c = c_B(\mu,k_B^{(i)}) = c_C(\mu,k_C^{(j)})  \ ,
\ee
then none of the four different algebras we obtain from $\WeB[\mu]$ for $k_B=k_B^{(i)}$ and 
$\WeC[\mu]$ for $k_C=k_C^{(j)}$ are equivalent to one another. Thus there is not a `unique' quantisation
of $\We[\mu]$! 

The two constructions are, however, closely related to one another since we have the identifications
\begin{equation}\label{eq:langlands}
\begin{split}
c_B(\mu+1,k_B)&=c_C(\mu,k_C)\\
\gamma_B(\mu+1,k_B) &= \gamma_C(\mu,k_C)
\end{split} \qquad \text{when}\quad
(k_B+\mu-1)(k_C+\mu+2) =1\ .
\end{equation}
This relation is  the `analytic continuation' of the Langlands duality that relates 
$B_n=\mathfrak{so}(2n+1)$ and $C_n=\mathfrak{sp}(2n)$. Indeed, the Dynkin diagrams of
$B_n$ and $C_n$ are obtained from one another upon reversing the arrows, i.\,e.\ upon exchanging the
roles of the long and the short roots. Correspondingly, the root system of one algebra can be identified
with the coroot system of the other (provided we scale the roots and coroots appropriately --- this is the
reason for our non-standard normalisation convention for the roots of $C_n$). It is then manifest from
the above formulae that the central charge and spectrum is the same provided we also exchange the roles
of $\alpha_+$ and $\alpha_-$. In terms of the levels $k_B$ and $k_C$, this is then equivalent to the requirement
that $(k_B+\mu-1)(k_C+\mu+2) =1$  for $\mu=2n$. Thus we can think of $\WeB[\mu+1]$ and 
$\WeC[\mu]$ to be related by Langlands duality for all $\mu$. 

The ambiguity in the definition of the quantum algebra associated with $\We[\mu]$ therefore 
simply reflects
that Langlands duality acts non-trivially on $\hse$, i.\,e.\ that $\hse$ is non-simply-laced. 
This is to be contrasted with the case of 
${\cal W}_{\infty}[\mu]$ where the two solutions of $k$ 
for a given $\mu$ and $c$ actually gave rise to equivalent ${\cal W}_\infty$ algebras, see eq.~(2.9) 
of \cite{Gaberdiel:2012ku}, reflecting
the fact that ${\rm hs}[\mu]$ can be thought of as being `simply-laced'.


\subsection{Classical limit}
\label{sec:cllim}

In the semiclassical limit of large levels, the two quantum algebras $\WeB[\mu]$ and $\WeC[\mu]$ 
actually become equivalent. More concretely, if we choose the 
normalisation of $n_4$ as in section~\ref{sec:wedge}, we have in the semiclassical limit
\begin{align}\label{eq:cl_lim}
c_B&\sim -\tfrac{1}{2} \mu(\mu^2-1) k_B+\mathcal{O}(k_B^0)\ ,& 
c_C&\sim -\tfrac{1}{2}\mu(\mu^2-1)k_C +\mathcal{O}(k_C^0)\  ,\\
\gamma_B & \sim \tfrac{144}{5}(\mu^2-19)^2+\mathcal{O}(k_B^{-1})\ ,& 
\gamma_C & \sim \tfrac{144}{5}(\mu^2-19)^2+\mathcal{O}(k_C^{-1}) \ .
\label{eq:glim}
\end{align}
In particular, the central charges agree, and the parameter $\gamma$ is of the form predicted by 
eq.~(\ref{eq:c444}), recalling that $\gamma=(c_{44}^{4})^2$. 
Thus both quantum algebras $\WeB[\mu]$ and $\WeC[\mu]$  define consistent quantisations of 
the classical Poisson algebra, and both can be thought of as Drinfel'd-Sokolov reductions of
$\hse$. However, as mentioned before, the $\mathcal{O}(c^{-1})$ corrections in eq.~\eqref{eq:glim} are 
different, reflecting that non-trivial action of Langlands duality as described by eq.~(\ref{eq:langlands}).


\subsection{Self-dualities}


While the parameters $\mu$ and $k$ are well suited for characterising the classical limits of the algebras 
$\WeB[\mu]$ and $\WeC[\mu]$, they do not directly parametrise the inequivalent $\We$ algebras. 
(The following discussion is directly parallel to the analogous analysis for the case of 
${\cal W}_{\infty}[\mu]$ in \cite{Gaberdiel:2012ku}.)
Indeed, as was stressed in section~\ref{eq:associativity}, the parameters distinguishing between 
different $\We$ algebras are $c$ and $\gamma$. It follows from 
eq.~(\ref{eq:cgbk}) that there are $12$ different combinations $(\mu_i,k_B^{(i)})$ that give rise to the
{\em same} quantum algebra $\WeB[\mu]$, and likewise for $\WeC[\mu]$, see eq.~(\ref{eq:cgck}).
Six of these identifications can be written down simply, while the other six require cubic roots; 
the simple identifications for $\WeB[\mu]$ relate $(\mu,k_B)$ to 
\begin{equation}\label{wbsd}
\begin{array}{l}
\bigl(\mu_2,k_B^{(2)}\bigr) = \bigl( \mu^2 + \mu (k_B -2)+k_B -1\ , \ 
3-\frac{1}{\mu+k_B -2}-  \mu_2 \bigr) \quad  \\[4pt]
\bigl(\mu_3,k_B^{(3)}\bigr) = \bigl( \mu ^2 + \mu (k_B -4) -k_B +4\ ,\  
\frac{1}{\mu+k_B -3}+3  -\mu_3  \bigr) \\[4pt]
\bigl(\mu_4,k_B^{(4)}\bigr) = \bigl(\frac{\mu  (\mu+k_B -3)}{\mu+k_B -2}\ ,\  
-\mu-k_B +5 -\mu_4  \bigr) \\[4pt]
\bigl(\mu_5,k_B^{(5)}\bigr) =\bigl( \frac{2}{\mu+k -3}+\mu +1  \ ,
 \frac{1}{\mu+k_B -2}+2 - \mu_5 \bigr) \\[4pt]
\bigl(\mu_6,k_B^{(6)}\bigr) =\bigl( -\frac{k_B}{\mu+k_B -2}-\mu +2 \ , \ 
2-\frac{1}{\mu+k_B -3}  - \mu_6 \bigr) \ .
\end{array}
\end{equation}
Note that all of these identifications are generated by the two primitive transformations
$(\mu,k_B)\mapsto (\mu_2,k_B^{(2)})$ and $(\mu,k_B)\mapsto (\mu_3,k_B^{(3)})$.
Similarly, for $\WeC[\mu]$ the simple identifications relate  $(\mu,k_C)$ to
\begin{equation}\label{wcsd}
\begin{array}{l}
\bigl(\mu_2,k_C^{(2)}\bigr) = \bigl( \mu^2 + \mu (k_C +2)+k_C \ , \ 
\frac{1}{\mu + k_C +1}-1 -\mu_2 \bigr) \\[4pt]
\bigl(\mu_3,k_C^{(3)}\bigr) = \bigl(  \mu^2 + \mu k_C - k_C -3\ ,\  
-\frac{1}{\mu+k_C  +2}-1 - \mu_3 \bigr) \\[4pt]
\bigl(\mu_4,k_C^{(4)}\bigr) = \bigl(  -\frac{\mu  (\mu + k_C + 1)}{\mu + k_C +2} \ , \
-\mu - k_C - 3 -\mu_4 \bigr) \\[4pt]
\bigl(\mu_5,k_C^{(5)}\bigr) =\bigl(  -\frac{2}{\mu+k_C  +1}+\mu -1 \ , \ 
\frac{1}{\mu+k_C  +2}-2 -\mu_5 \bigr) \\[4pt]
\bigl(\mu_6,k_C^{(6)}\bigr) =\bigl(  -\frac{k_C}{\mu+k_C  +2}-\mu \ , \
 -\frac{1}{\mu+k_C  +1}- 2-\mu_6 \bigr) \ .
\end{array}
\end{equation}
Again, all of these identifications are generated by the two primitive transformations
$(\mu,k_C)\mapsto (\mu_2,k_C^{(2)})$ and $(\mu,k_C)\mapsto (\mu_3,k_C^{(3)})$.

\section{The coset constructions}

It was proposed in \cite{Ahn:2011pv,Gaberdiel:2011nt} that the higher spin theory on AdS$_3$ based on 
the even spin algebra --- from what we have said above, it is now clear that the relevant
algebra is in fact ${\rm hs}^e[\lambda]$ --- should be dual to the 't~Hooft limit of the $\mathfrak{so}(2n)$ cosets
\begin{equation}\label{eq:mm_def}
{\cal WD}_{n,k} = \frac{\mathfrak{so}(2n)_k \oplus \mathfrak{so}(2n)_1}{\mathfrak{so}(2n)_{k+1}} \ ,
\end{equation}
where the 't~Hooft limit consists of taking $n,k\rightarrow \infty$ while keeping the parameter 
\begin{equation}\label{eq:limit}
\lambda = \frac{2n}{2n+k-2} \quad \text{fixed.}
\end{equation}
This therefore suggests that the corresponding quantum $\We$ algebras should be isomorphic. Given
that there are two different quantisations of the Drinfel'd-Sokolov reduction of $\hse$ (see section~3), 
there should therefore be two identifications, relating ${\cal WD}_{n,k}$ to either 
$\WeB[\lambda]$ or $\WeC[\lambda]$. In this section we want to explain in detail these different relations. 
As  in the case of ${\cal W}_{\infty}[\mu]$ studied in \cite{Gaberdiel:2012ku}, the (correctly adjusted) 
correspondences will actually turn out to hold even at finite $n$ and $k$.


\subsection{The \texorpdfstring{$D_n$}{Dn} cosets}
\label{sec:chalg}

In a first step we need to understand the structure of the ${\cal W}$ algebra underlying the
cosets (\ref{eq:mm_def}). By the usual formula we find that its central charge equals 
\begin{equation}\label{eq:cmm}
c_{{\mathfrak{so}}}(2n,k) = n \left[1-\frac{(2n-2) (2n-1)}{(k+2n-2) (k+2n-1)}\right] \ .
\end{equation}
In order to determine the spin spectrum of the ${\cal W}$ algebra we can use that $D_n$ is simply-laced,
and hence that (\ref{eq:mm_def}) is isomorphic \cite{Bouwknegt:1992wg}
to the Drinfel'd-Sokolov reduction of $D_n$, which we denote by ${\cal WD}_n$; this algebra is  
of type ${\cal W}(2,4,\ldots,2n-2,n)$. In the 't~Hooft limit, i.\,e.\ for 
$n\rightarrow \infty$, the spin spectrum of ${\cal WD}_n$ involves all even spins (with multiplicity one),
and hence becomes a $\We$ algebra, but for finite $n$, this is not the case because of the additional
spin $n$ generator, which we shall denote by $V$. However, as was already explained 
in \cite{Honecker:1992kz,Blumenhagen:1994wg},
${\cal WD}_n$ possesses an outer $\mathbb{Z}_2$ automorphism $\sigma$ 
--- this is actually the automorphism that is 
inherited from the spin-flip automorphism of $\mathfrak{so}(2n)$ --- under which the generators of spin 
$2,4,\ldots, 2n-2$ are invariant, while the spin $n$ generator $V$ is odd.
Then, the `orbifold' subalgebra 
${\cal WD}^\sigma_n$, i.\,e.\ the $\sigma$-invariant subalgebra of ${\cal WD}_n$, has the right spin content.
It is generated, in addition to the $\sigma$-invariant generators of $\mathcal{WD}_n$ of spin $2,4,\ldots,2n-2$, by 
the normal ordered product of spin $2n$ $:V V:$, as well as its higher derivatives 
that are schematically of the form $: V \partial^{2l} V:$, see \cite{Blumenhagen:1994wg}.\footnote{The 
counting of the quasiprimary higher spin states is essentially equivalent to the counting of the 
higher spin fields of a theory of a single real boson, see e.\,g.\ \cite{Klebanov:2002ja}.
Note that, as is also explained in 
\cite{Blumenhagen:1994wg}, the resulting algebra is neither freely generated nor infinitely generated, i.\,e.\ there are relations between the $\We$
type generators that effectively reduce these generators to a finite set.}

These arguments imply that we can generate ${\cal WD}^\sigma_n$  by (a subset of) the fields contained in $\We$.
Hence,  ${\cal WD}^\sigma_n$ is a quotient of $\We$ and we can characterise it again in terms of the central
charge $c$, and the  parameter $\gamma=(c_{44}^{4})^2$. As before, a convenient method to compute
$\gamma$ is by comparing the conformal dimension of the `minimal' representations.
Since ${\cal WD}^\sigma_n$ is a subalgebra of ${\cal WD}_n$, each representation of 
${\cal WD}_n$ defines also a representation of ${\cal WD}^\sigma_n$. In particular, the `minimal'
representations of ${\cal WD}_n$ that are labelled by $(v;0)$ and $(0;v)$ --- see  \cite{Gaberdiel:2011nt} 
for our conventions --- are also minimal for ${\cal WD}^\sigma_n$, and their conformal dimensions equal 
\begin{equation}\label{eq:hsomin}
h(v;0) = \frac{1}{2} \left[ 1 + \frac{2n-1}{k+2n-2} \right] \ ,\qquad  \ 
h(0;v) = \frac{1}{2} \left[1-\frac{2n-1}{k+2n-1}\right] \ .
\end{equation}
Both solve eq.~\eqref{eqn:gamma0} for $\gamma=\gamma_{\mathfrak{so}}(N,k)$, where $N=2n$ and 
\begin{align} 
\gamma_{\mathfrak{so}} =&\ 144 (-224 + 744 k - 860 k^2 + 408 k^3 - 68 k^4 + 376 N - 1336 k N + 1267 k^2 N \notag \\
& - 386 k^3 N  + 19 k^4 N - 124 N^2 + 857 k N^2 - 599 k^2 N^2 + 76 k^3 N^2 - 52 N^3 \notag \\
& - 252 k N^3 + 94 k^2 N^3
 + 24 N^4 +  36 k N^4)^2 n_4\, /\, \Bigl[
c_{\mathfrak{so}} (2 + k) (N-1) (2 k-4 + N) \notag  \\ 
& \times (k-5 + 2 N) (3 k-4 + 2 N)
(2 k-2 + 3 N)(3 k-5 +      4 N)  \notag \\
& \times (88 - 132 k + 44 k^2 - 132 N + 73 k N + 5 k^2 N+ 44 N^2 + 10 k N^2) \Bigr]  \ .
\label{eq:gmm}
\end{align}
It is interesting that also $h=n$ solves eq.~\eqref{eqn:gamma0} for $\gamma=\gamma_{\mathfrak{so}}(2n,k)$, thus
implying that also the field $V$ generates a minimal representation of ${\cal WD}^\sigma_n$.


\subsection{The \texorpdfstring{$B_n$}{Bn} cosets}\label{oddcos}


A closely related family of cosets is obtained from (\ref{eq:mm_def}) by considering instead
the odd $\mathfrak{so}$ algebras, i.\,e.\
\begin{equation}\label{eq:oddcosets}
\mathcal{WB}(0,n)^{(0)} = \frac{\mathfrak{so}(2n+1)_k \oplus \mathfrak{so}(2n+1)_1}{\mathfrak{so}(2n+1)_{k+1}} \ .
\end{equation}
These ${\cal W}$ algebras can be identified with the bosonic subalgebra of the Drinfel'd-Sokolov reduction of
the superalgebras $\mathfrak{osp}(1|2n)$ or $B(0,n)$, see~\cite{Bouwknegt:1992wg}. The latter is a 
${\cal W}$ algebra of type 
$\mathcal{W}(2,4,\dots, 2n,n+\tfrac{1}{2})$, and we shall denote it by $\mathcal{WB}(0,n)$. Since
the field of conformal weight $n+\tfrac{1}{2}$ is fermionic --- we shall denote it by $S$ in the following --- 
the bosonic subalgebra does not include $S$,
but contains  instead
the normal ordered products  $:S\partial^{2l+1} S:$ with $l=0,1,\ldots$ --- because $:SS:=0$ we 
now always have an odd number of derivatives.\footnote{The counting of the quasiprimary higher spin
fields is in this case analogous to that of counting the higher spin fields of a single free fermion.}
Thus the generating fields include, in addition to the 
bosonic generating fields of $\mathcal{WB}(0,n)$ of spin $2,4,\ldots, 2n$, fields
of spin $2n+2,2n+4,\ldots$; in particular, $\mathcal{WB}(0,n)^{(0)}$ is therefore again
a quotient of $\We$, and can be characterised in terms of $\gamma$ and $c$.
The analysis is 
essentially identical to what was done for the $\mathfrak{so}(2n)$ case above --- indeed, the central
charge, as well as the conformal dimensions of the minimal representations are obtained from 
(\ref{eq:cmm}) and (\ref{eq:hsomin}) upon replacing $2n\mapsto 2n+1$, and thus 
$\gamma$ is simply $\gamma=\gamma_{\mathfrak{so}}(N,k)$, where $N=2n+1$ and 
$\gamma_{\mathfrak{so}}$ was  already defined in (\ref{eq:gmm}). Thus these two families of cosets are 
naturally analytic continuations of one another.

As an additional consistency check we note that the 
algebra $\mathcal{WB}(0,1)^{(0)}$ is of type $\mathcal{W}(2,4,6)$, see~\cite{Bouwknegt:1988nz}, and its structure 
constants are explicitly known \cite{Honecker:1992nr}.
In section~\ref{sec:w246} we have reproduced this algebra as a quotient of $\We$.
The corresponding value of $\gamma$, given in eq.~\eqref{eq:gamma_set1},  agrees indeed 
with $\gamma_{\mathfrak{so}(3)}$.


\subsection{Level-rank duality}


The expressions~(\ref{eq:cmm}) and (\ref{eq:gmm}) are invariant under the transformation
\begin{equation}\label{eq:mmkl}
N\mapsto N\ ,\qquad k\mapsto-2N-k+3\ .
\end{equation}
For even $N=2n$ this is a consequence of the  Langlands self-duality of $D_n$, which in turn follows from the 
fact that $D_n$ is simply-laced, implying that the Drinfel'd-Sokolov reduction has the 
symmetry $\alpha_\pm\mapsto -\alpha_\mp$. As a result, ${\cal WD}_n$ actually only depends on
$c$, rather than directly on $k$. This is reflected in the fact that $\gamma_{\mathfrak{so}}$ can be written 
as  an unambiguous function of $N$ and $c$ as 
\begin{align}\notag
\gamma_{\mathfrak{so}} &= 72 \bigl(2 c^2 (N^2-2N-18)+3 c (6 N^3-49 N^2+80 N-8)+2N (6N^2 + 5N -28)\bigr)^2 \notag \\
& \quad{}\times n_4/ \Bigl[
(5 c +22)c (c (N^2-7N+12)+2 N^2-5N) (c (N+1) + 4 N^2-5N ) \notag \\
& \qquad \quad \times (2 c (N+2) +3N^2-14N+8) \Bigr]\ .
   \label{eqn:gammaD}
\end{align}
Note that, in the large $c$ limit, $\mathcal{WD}^\sigma_n$ becomes a classical Poisson 
algebra, which can be identified with the $\sigma$-invariant classical Drinfel'd-Sokolov reduction of $D_n$.
In fact, taking $n_4$ as in eq.~\eqref{eq:n4}, it follows from eq.~(\ref{eqn:gammaD}) that the corresponding
$\gamma$ parameter equals
\begin{equation}\label{eq:mudef}
\gamma_{\mathfrak{so}} = \tfrac{144}{5}(\mu^2-19)^2+\mathcal{O}(c^{-1})\ ,\qquad \text{where}\quad \mu=2n-1\ .
\end{equation}
Note that this ties in with the fact that the 
wedge algebra of $\mathcal{WD}^\sigma_n$ is the $\sigma$-invariant subalgebra of $\mathfrak{so}(2n)$, 
which in turn equals $\mathfrak{so}(2n-1)$. This explains why (\ref{eq:mudef}) agrees with (\ref{eq:glim}) for $\mu=2n-1$. 
\smallskip

Next we observe that eq.~\eqref{eqn:gammaD} is a polynomial equation of order $6$ in $N$, with
coefficients that are functions of $\gamma$ and $c$, and hence there is a six-fold ambiguity in 
the definition of $N$. If we parametrise $c=c_{\mathfrak{so}}(N,k)$, then the algebra associated
with $(N,k)$ is equivalent to the one associated with 
\begin{equation}\label{mmsd}
(N_2,k_2)=\Bigl( \frac{k+2N-3}{k+N-2}\ , \ \frac{k}{k+N-2}  \Bigr) \ , \quad 
(N_3,k_3)= \Bigl( \frac{k}{k+N-1} \ , \ \frac{2N+k-3}{k+N-1}\Bigr) \ ,
\end{equation}
while the other three solutions involve cubic roots. Obviously, we can also replace 
$k\mapsto -2N-k+3$ without modifying the algebra, see (\ref{eq:mmkl}), and thus, expressed 
in terms of $N$ and $k$, there are $12$ different pairs $(N_i,k_i)$ that define the same algebra.
We should also mention that the third solution above is obtained by applying the map
$(N,k)\mapsto (N_2,k_2)$ twice. This fundamental transformation has a nice interpretation
in terms of a level-rank type duality rather similar to the one 
appearing for $\mathfrak{su}(N)$ in \cite{Gaberdiel:2012ku}:
\begin{equation}\label{eq:mm_lrd}
\left(\frac{\mathfrak{so}(N)_{k} \oplus \mathfrak{so}(N)_1} {\mathfrak{so}(N)_{k + 1}}\right)^\sigma \cong \left(\frac{\mathfrak{so}(M)_{l} \oplus \mathfrak{so}(M)_1} {\mathfrak{so}(M)_{l + 1}}\right)^\sigma  \ ,
\end{equation}
where
\begin{equation}\label{eqn:levelRank}
k=\frac{N-1}{M-1} - N + 2 \ , \qquad \ l=\frac{M - 1}{N - 1} - M + 2\ ,
\end{equation}
and the superscript $\sigma$ means that we take the $\sigma$-invariant subalgebra if $N$ or $M$ are
even integers. Obviously, as a true level-rank duality, this only 
makes sense if $M$ and $N$ are positive
integers. As far as we are aware, this level-rank duality has not been noticed before.


\subsection{Holography}


With these preparations we can now return to the main topic of this section, the precise relation
between the $\sigma$-even subalgebra of the $\mathfrak{so}(2n)$ cosets of eq.~(\ref{eq:mm_def}), and the quantum
algebras $\WeB[\mu]$ and $\WeC[\mu]$. As we have explained before, all three algebras are
in general (quotients of) $\We$ algebras, and hence are uniquely characterised in terms of $\gamma$ and $c$. 
By comparing the relations (\ref{eq:redcB}) and (\ref{eq:cgbk}) for $\WeB[\mu]$ with 
(\ref{eq:cmm}) and (\ref{eq:gmm}) for the $\mathfrak{so}(2n)$ cosets, we conclude that we have the identification
\begin{equation}\label{eq:dbhol}
{\cal WD}_{n,k}^{\sigma} \cong 
\WeB[\lambda_B] \ , \quad \hbox{with} \quad
\lambda_B = \frac{2n-2}{k+2n-2} \ , \quad k_B  = k + 2n + 1-\lambda_B \ . 
\end{equation}
Similarly, for the case of $\WeC[\mu]$ we find instead from (\ref{cC}) and (\ref{eq:cgck}) that
\begin{equation}\label{eq:dchol}
{\cal WD}_{n,k}^{\sigma} \cong 
\WeC[\lambda_C] \ , \quad \hbox{with} \quad
\lambda_C = \frac{2n}{k+2n-2} \ , \quad k_C  = k + 2n - 3-\lambda_C \ . 
\end{equation}
Obviously, using the self-duality relations of the various algebras, see 
eqs.~(\ref{wbsd}), (\ref{wcsd}) and (\ref{mmsd}), there 
are also other versions of these identifications, but the above is what is relevant
in the context of minimal model holography: the above analysis shows that the ($\sigma$-even subalgebra
of the) $\mathfrak{so}$ cosets\footnote{For $n\in \mathbb{N}+\tfrac{1}{2}$, the left hand side of 
eqs.~(\ref{eq:dbhol}) and (\ref{eq:dchol}) should be understood as the chiral algebra of the 
cosets~\eqref{eq:oddcosets}.
Coset interpretations exist also when $n$ is a negative half-integer, see section~\ref{sec:other_mm}.} 
are equivalent to the quantum Drinfel'd-Sokolov reduction of the
${\rm hs}^{e}[\lambda_{B/C}]$ algebras with $\lambda_{B/C}$ given above. Note that 
$\lambda_C$ agrees exactly with $\lambda$ given in (\ref{eq:limit}) above, see also \cite{Gaberdiel:2011nt},
while for $\lambda_B$ the difference is immaterial in the 't~Hooft limit. These statements are now
true even at finite $n$ and $k$, hence giving the correct 
quantum version of the even spin holography conjecture.


\subsection{The semiclassical behaviour of the scalar fields}


With our detailed understanding of the symmetry algebras at finite $c$, we can now also address
the question of whether the duals of the two minimal coset fields of \cite{Ahn:2011pv,Gaberdiel:2011nt} 
should be thought of as being perturbative or non-perturbative excitations of the higher spin bulk theory. 
As in the case studied in \cite{Gaberdiel:2012ku}, this issue can be decided by studying the behaviour 
of their conformal dimensions in the semiclassical limit, i.\,e.\ for $c\rightarrow \infty$. 

Let us consider then the $\mathcal{WD}^\sigma_n$ coset at fixed $n$. If $c$ takes one of the
actual minimal model values, $c=c_{\mathfrak{so}}(2n,k)$ with $k\in\mathbb{N}$, see
eq.~(\ref{eq:cmm}), the algebra has the two minimal representations $(v;0)$ and $(0;v)$, whose 
conformal dimensions are given in eq.~\eqref{eq:hsomin}. Written in terms of $n$ and $c$ (rather than $n$ and $k$),
they take the form 
\begin{equation}\label{hpm}
h_{\pm}(n,c) = \frac{1}{2} \left( 1+\frac{n -c \pm \sqrt{(c-n) \left(c-(3-4 n)^2 n\right)}}{4 (n-1) n}\right) \ ,
\end{equation}
where $h(v;0)=h_+(n,c)$ and $h(0;v)=h_{-}(n,c)$. Since we know that the algebra $\mathcal{WD}^\sigma_n$ depends
only on $c$ (rather than $k$), it is then clear that (\ref{hpm}) are the conformal weights of minimal representations
for any value of $c$. 

We are interested in the semiclassical limit, which consists of taking  $c\rightarrow \infty$ at fixed $n$. There is obviously 
an ambiguity in how precisely $c$ is analytically continued, but taking $c$, say, along the positive real axis to infinity, 
we read off from (\ref{hpm}) that 
\begin{align}
h(v;0) &= h_{+}(n,c) = \frac{1-\mu}{2} +\mathcal{O}(c^{-1})\ , \\ 
h(0;v) &= h_{-}(n,c) = \frac{c}{\mu^2-1}  + \mathcal{O}(1)\ ,
\end{align}
where $\mu=2n-1$, see eq.~\eqref{eq:mudef}. In this limit $h(v;0)$ remains finite, while $h(0;v)$ is proportional
to $c$. Thus we conclude that only the coset representation  $(v;0)$ corresponds to a perturbative scalar of 
the higher spin theory based on $\hse$, while  $(0;v)$ describes a non-perturbative excitation. This is directly analogous to
what happened in \cite{Gaberdiel:2012ku}.


\subsection{The full orbifold spectrum}\label{sec:full}


Now that we have understood the relation between the symmetries in the 
duality conjecture of \cite{Ahn:2011pv,Gaberdiel:2011nt} we can come
back to the comparison of the partition functions that was performed in 
\cite{Gaberdiel:2011nt}. It was shown there that the spectrum of the charge 
conjugate modular invariant of the ${\cal WD}_{n,k}$ algebra coincides, in
the 't~Hooft limit, with the bulk 1-loop partition function of a suitable higher spin theory 
on thermal AdS$_3$. 

As we have seen above, at finite $n$ and $c$, the relevant symmetry algebra is 
actually not ${\cal WD}_{n,k}$, but only the $\sigma$-invariant subalgebra 
${\cal WD}_{n,k}^\sigma$. Every representation of ${\cal WD}_{n,k}$ defines
also a representation of ${\cal WD}_{n,k}^\sigma$, and hence the 
charge conjugation (or A-type)  modular invariant of the ${\cal WD}_{n,k}$ algebra
also defines a consistent partition function with respect to ${\cal WD}_{n,k}^\sigma$.
However, from the latter point of view, it is not the charge conjugation modular invariant,
but rather of what one may call `D-type'.

It is then natural to ask whether the charge conjugation (A-type) modular invariant of ${\cal WD}_{n,k}^\sigma$
also has a bulk interpretation. We shall not attempt to answer this question here, but 
we shall only show that it leads to a {\em different} partition function in the 't~Hooft limit. Thus, if the charge-conjugation
modular invariant of  ${\cal WD}_{n,k}^\sigma$ also has
a consistent AdS$_3$ dual, this must be a different theory than the one considered in 
\cite{Ahn:2011pv,Gaberdiel:2011nt}}.
\smallskip

In the charge conjugation (A-type) modular invariant of the ${\cal WD}_{n,k}^\sigma$ algebra,
every untwisted representation of ${\cal WD}_{n,k}^\sigma$ appears once. Obviously, not all representations of 
${\cal WD}_{n,k}^\sigma$ arise as subrepresentations of untwisted ${\cal WD}_{n,k}$ representations. In particular,
each $\sigma$-twisted representation of ${\cal WD}_{n,k}$ (for which $V$ is half-integer moded) also leads to 
an untwisted representation of ${\cal WD}_{n,k}^\sigma$. Since $\sigma$ is inherited from the 
outer automorphism of $\mathfrak{so}(2n)$, these twisted representations of $\mathcal{WD}_{n,k}$
can be described via the cosets
\begin{equation}\label{eq:twcosets}
\frac{\mathfrak{so}(2n)^{(2)}_k\oplus \mathfrak{so}(2n)^{(2)}_1}{\mathfrak{so}(2n)^{(2)}_{k+1}}\ ,
\end{equation}
where $\mathfrak{so}(2n)^{(2)}_k$ is the twisted affine algebra, see e.\,g.\ \cite{Goddard:1986bp} for an 
introduction. The  representations of $\mathfrak{so}(2n)^{(2)}_k$ are labelled by $\mathfrak{so}(2n-1)$ 
dominant highest weights $\Xi$, satisfying certain integrability conditions, and the corresponding
conformal dimensions equal
\begin{equation}\label{eq:twis}
h_{\mathfrak{so}(2n)_k^{(2)}}(\Xi)=
\frac{\mathrm{Cas} (\Xi)}{2(k+2n-2)} + \frac{k(2n-1)}{16(k+2n-2)}\ ,
\end{equation}
where ${\rm Cas}$ is the Casimir of $\mathfrak{so}(2n-1)$.
The conformal dimension of the representations of \eqref{eq:twcosets} can then be obtained from 
\eqref{eq:twis} by the usual coset formula.
In particular, the twisted vacuum, where we take $\Xi$ to be the vacuum representation ($\Xi=0$) of 
$\mathfrak{so}(2n-1)$ for all $3$ factors in eq.~\eqref{eq:twcosets},  has conformal dimension 
\begin{equation}
\frac{1}{16}\left[ 1- \frac{(2n-1)(2n-2)}{(k+2n-2)(k+2n-1)} \right]\ .
\end{equation}
This state does not appear in the $1$-loop bulk higher spin calculation of \cite{Gaberdiel:2011nt}, and thus the dual
of the charge conjugation modular invariant of ${\cal WD}_{n,k}^\sigma$ must be a different bulk theory
than the one considered in \cite{Gaberdiel:2011nt}.


\subsection{Other minimal models}\label{sec:other_mm}


Let us close this discussion with a comment about other minimal models one may consider. As we have seen in 
sections~\ref{sec:chalg} and \ref{oddcos},  the dual of the
even higher spin theories on AdS can be identified with the cosets of {\em either} the 
$\mathfrak{so}(\hbox{even})$ or the  $\mathfrak{so}(\hbox{odd})$ algebras. It is then
natural to ask how  the cosets of the $\mathfrak{sp}$ algebras fit into this picture.
Using the field counting techniques of~\cite{Candu:2012jq} (see also \cite{Creutzig:2012ar}) one can 
show that the cosets\footnote{In our conventions, the {\em short} roots of $\mathfrak{sp}(2n)$ have  
length squared equal to $2$.}
\begin{equation} \label{eq:spcoset}
\frac{\mathfrak{sp}(2n)_k \oplus \mathfrak{sp}(2n)_{-1}}{\mathfrak{sp}(2n)_{k -1}}
\end{equation}
possess a $\We$ symmetry in the 't~Hooft limit. The essential points of this calculation are (i) that 
$\mathfrak{sp}(2n)_{-1}$ has a free field realisation in terms of $n$ $\beta\gamma$-systems; and (ii) 
that the coset vacuum character can be computed by counting  $\mathfrak{sp}(2n)$ invariant products of 
$\beta\gamma$-fields and their derivatives, using standard arguments of classical invariant theory.

It is then natural to ask what $\We$ algebras the cosets~\eqref{eq:spcoset} lead to 
when analytically continued in $n$ and $k$. 
The answer can be schematically formulated as
\begin{equation}\label{ident}
\frac{\mathfrak{sp}(2n)_k \oplus \mathfrak{sp}(2n)_{-1}}{\mathfrak{sp}(2n)_{k -1}} \cong \left(
\frac{\mathfrak{so}(-2n)_{-k} \oplus \mathfrak{so}(-2 n)_{1}}{\mathfrak{so}(-2n)_{-k +1}} \right)^\sigma\ ,
\end{equation}
where both cosets stand for the corresponding $\We$ algebras (or their quotients), and the equality means that
both the analytically continued central charge and the self-coupling $\gamma$ agree.

Incidentally, there is an independent check for our claim that the cosets \eqref{eq:spcoset} 
are quotients of $\We$. For $n=1$, the coset \eqref{eq:spcoset} 
is known to be of  type $\mathcal{W}(2,4,6)$, see~\cite{deBoer:1993gd},\footnote{We 
thank the authors of \cite{Creutzig:2012ar} for drawing our attention
to this reference.}
and its structure constants have been computed explicitly  in \cite{Eholzer:1993ak}, coinciding 
with the solution given in eq.~\eqref{eq:gamma_set2} of section~\ref{sec:w246}. We also note that the 
corresponding value of $\gamma$ agrees indeed with 
$\gamma_{\mathfrak{so}(-2)}$, as required by (\ref{ident}). 

The above arguments apply similarly for the cosets
\begin{equation}
\frac{\mathfrak{osp}(1|2n)_k \oplus \mathfrak{osp}(1|2n)_{-1}}{\mathfrak{osp}(1|2n)_{k -1}}\ ,
\end{equation}
for which the emergence of a $\We$ symmetry in the 't~Hooft limit can be proven using analogous methods, 
in particular, noting that $\mathfrak{osp}(1|2n)_{-1}$
has a free field realisation in terms of a single Majorana fermion and $n$ $\beta\gamma$-systems.
In this case, the analogue of (\ref{ident}) is 
\begin{equation}
\frac{\mathfrak{osp}(1|2n)_k \oplus \mathfrak{osp}(1|2n)_{-1}}{\mathfrak{osp}(1|2n)_{k -1}}  \cong
\frac{\mathfrak{so}(-2n+1)_{-k} \oplus \mathfrak{so}(-2 n+1)_{1}}{\mathfrak{so}(-2n+1)_{-k +1}}
 \ .
\end{equation}
%


\section{Conclusions}


In this paper we have constructed the quantum $\We$ algebra that is generated by one Virasoro
primary field for every even spin, using systematically Jacobi identities. We have seen that, up to the level
to which we have evaluated these constraints, the algebra depends only
on two parameters: the central charge $c$ and a free parameter $\gamma$, which is
essentially the self-coupling of the spin 4 field. We have shown that the first few commutators of the 
wedge algebra of $\We$ agree with those of $\hse$. This suggests that the dual higher spin theory
on AdS$_3$ should be described in terms of a Chern-Simons theory based on $\hse$. 
Furthermore, given the usual relation between
wedge algebras and Drinfel'd-Sokolov reductions, $\We$ should be thought of as the 
quantum Drinfel'd-Sokolov reduction of $\hse$. As we have explained, there are actually
two different quantisations of the classical Drinfel'd-Sokolov reduction of $\hse$, which we called $\WeB$
and $\WeC$, respectively.  We have argued that this ambiguity is closely related to the fact that $\hse$ is 
non-simply-laced.

Given that $\We$ describes the most general $\W$ algebra with this spin content, we can
identify (quotients of) $\We$  with (orbifolds of) the coset algebras based on 
$\mathfrak{so}(2n)$, $\mathfrak{so}(2n+1)$, $\mathfrak{sp}(2n)$, and $\mathfrak{osp}(1|2n)$.
In particular, this proves that the $\mathfrak{so}$ coset algebras of \cite{Ahn:2011pv,Gaberdiel:2011nt}
are equivalent, for suitable values of $\mu$, to the quantum algebras $\WeB[\mu]$ and $\WeC[\mu]$.
This quantum equivalence is  even true at finite $n$ and $k$, and therefore
establishes an important part of the holographic proposals of \cite{Ahn:2011pv,Gaberdiel:2011nt}. 
We also showed, in close analogy with \cite{Gaberdiel:2012ku}, that only one of the `scalar' excitations 
should be thought of as being perturbative, while the other should correspond to a non-perturbative
classical solution. It would be interesting to check, following \cite{Castro:2011iw,Perlmutter:2012ds},
whether the corresponding classical solutions exist and have the appropriate properties. It would
also be interesting to study whether the A-type modular invariant of the ${\cal WD}_{n,k}^\sigma$ algebra
has a bulk interpretation, see section~\ref{sec:full}. 

As in the case of the higher spin algebra $\hs{\mu}$ discussed in 
\cite{Gaberdiel:2012ku}, it would be interesting to reproduce the quantum corrections predicted by the 
CFT directly from a perturbative bulk calculation.  In this context it would be important 
to understand the systematics of the quantum Drinfel'd-Sokolov reduction for $\hse$ in more detail. 
In particular, this should shed some light on which choices have to be made in
quantising the bulk theory.

\section*{Acknowledgements} 
We thank Stefan Fredenhagen and Rajesh Gopakumar for useful conversations. The work of 
MRG, MK and CV is supported in part by the Swiss National Science Foundation. We thank
the ESI in Vienna for hospitality during an early stage of this work.

\appendix


\section{Minimal representations using commutators}
\label{sec:commutators}


In section~\ref{sec:min} we computed the structure constant $c_{44}^4$
in terms of the conformal dimension $h$ of minimal representations. The calculation was
carried out using OPEs. An alternative, but equivalent approach uses commutators rather than OPEs
and shall be sketched in this appendix.

\noindent
We will need the following commutators of $\mathcal{W}_{\infty}^{e}$:
\begin{align}
[L_m,L_n] = &\  (m-n) L_{m+n}+\tfrac{c}{12}m(m^2-1)\delta_{0,m+n}\ ,\notag\\ 
[L_m,W^4_n] = &\ ( 3 m-n) W^4_{m+n} \notag\\
[W^4_m,W^4_n] = &\  \frac{1}{2}(m-n)\left( c_{44}^6 W^6_{m+n}+ q_{44}^{6,1} Q^{6,1}_{m+n}
+ q_{44}^{6,2} Q^{6,2}_{m+n}+q_{44}^{6,3} Q^{6,3}_{m+n}\right) \notag \\       
&{}+\frac{1}{36} \left(m^2-m n+n^2-7\right) (m-n)\left(c_{44}^4 W^4_{m+n} + q_{44}^4 
  Q^4_{m+n} \right) \notag \\ 
&+  \bigl(
3(m^4+n^4) -(2mn+39)(m^2+n^2)+4m^2 n^2 +20 mn +108\bigr)\notag\\
&\quad{}\times\frac{1}{3360}(m-n) q_{44}^2 L_{m+n} \notag \\
&+ \frac{1}{5040}m (m^2-1)(m^2-4)(m^2-9) n_4 
   \delta _{0,m+n} \ ,
\end{align}
where the composite quasiprimary fields $Q^4,Q^{6,1},Q^{6,2}$ and $Q^{6,3}$ are given by
\begin{align}
Q^4 &= LL -\frac{3 L''}{10} \ , & Q^{6,1} &= L W^4 -\frac{{W^4}''}{6} \ , \notag\\
Q^{6,2} &=  L' L'-\frac{4}{5}  L'' L -\frac{\partial^4 L}{42} \ , &
Q^{6,3} &=
   L(LL) -\frac{1}{3}
   L' L' -\frac{19}{30}
   L'' L -\frac{\partial^4 L}{36} \ .
\end{align}
Solving the Jacobi identity $[L_m,[W^4_n,W^4_l]] + \mathrm{cycl.} = 0$, we find that
\begin{align} \label{eq:qpStrConst}
q_{44}^2 &= \frac{8 }{c} n_4\ ,\quad q_{44}^{6,1} = \frac{28 }{3 (c+24)} c_{44}^4 \ , &
q_{44}^{6,2} &= -\frac{2 (19c -524) }{
 3 c (2 c-1) (7 c+68)} n_4 \ , \notag \\
q_{44}^{6,3} &= \frac{96 (72 c+13) }{c (2 c-1) (5 c+22) (7 c+68)} n_4 \ , &
q_{44}^4 &= \frac{168 }{c (5 c+22)} n_4 \ . 
\end{align}
This fixes the structure constants of the Virasoro descendants in 
terms of their primaries. Similarly, by
considering Jacobi identities of higher level, we can reobtain 
in this manner the relations between structure constants given in section~\ref{eq:associativity}.

Recall from section~\ref{sec:min} that the defining property of a minimal representation is a character of the form
\be
\frac{q^h}{1-q} \prod_{s \in 2\mathbb{N}}^{\infty} \prod_{n=s}^{\infty} \frac{1}{1-q^n} = q^h (1 + q + 2 q^2 + 3 q^3 + \dots ) \ ,
\ee
where $h$ is the conformal dimension of the highest weight state $\Phi$.

Thus, at level $1$ all the states must be  proportional to $L_{-1}\Phi$, at level $2$ they are linear combinations of, say, 
 $L_{-1}^2 \Phi$ and $L_{-2} \Phi$, and at level $3$ of, for instance,  $L_{-3} \Phi$, $L_{-2}L_{-1} \Phi$ and $L_{-1}^3 \Phi$.
Therefore, we can  conclude that the representation must have null relations of the form
\begin{align}
 \mathcal{N}_{1W^4}&= (W^4_{-1}-\frac{2w^4}{h} L_{-1})\Phi\ ,\\
 \mathcal{N}_{2W^4}&= (W^4_{-2}+a L_{-1}^2+b L_{-2})\Phi\ ,\\
 \mathcal{N}_{3W^4}&= (W^4_{-3}+d L_{-3}+e L_{-2}L_{-1}+f L_{-1}^3 )\Phi \ ,
\end{align}
where  $w^4$ is the eigenvalue of the zero mode of $W^4$ on $\Phi$.
The coefficient in front of $L_{-1}$ in $\mathcal{N}_{1W^4}$ follows from the condition
\be
L_1 \mathcal{N}_{1 W^4}=0 \ .
\ee
Similarly, the coefficients $a$ and $b$ in $\mathcal{N}_{2W^4}$ can be determined from the conditions 
\be
L_{1}^{2} \mathcal{N}_{2 W^4}=0 \quad \mathrm{and} \quad
 L_{2} \mathcal{N}_{2 W^4}=0 \ ,
\ee
and $d$, $e$ 
and $f$ from $L_3 \mathcal{N}_{3 W^4}=0$, $L_2 L_1 \mathcal{N}_{3 W^4}=0$ and $L_1^{3} \mathcal{N}_{3 W^4}=0$.
The result is 
\begin{align}\notag
a&=-\frac{(5 c+16 h)w^4}{h (2 c h+c+2 h (8 h-5))} \ , \qquad b=\frac{4 (11-8 h) w^4}{2 c h+c+2 h (8 h-5)}\ , \\
d&=-\frac{6 \bigl[c (h+3) (2 h-3)+2h (h-2) (8 h-21)-22\bigr]w^4}{\bigl[(c-7) h+c+3 h^2+2\bigr]
 \bigl[2 c h+c+2 h (8 h-5)\bigr]} \ , \notag \\
e&=-\frac{12  \bigl[c (6h (h-1)-2)+h (h (8 h-15)+9)\bigr]w^4}{h \bigl[(c-7) h+c+3 h^2+2\bigr] 
\bigl[2 c h+c+2 h (8 h-5)\bigr]} \ , \notag \\
f&=-\frac{(5 c+22) (c-h)w^4}{h \bigl[(c-7)
   h+c+3 h^2+2\bigr] \bigl[2 c h+c+2 h (8 h-5)\bigr]}\ .
\end{align} 
Finally,  solving the slightly more involved null relations 
\begin{equation}\label{eqn:null}
 W^4_1 \mathcal{N}_{1W^4} = 0\ ,\qquad
 W^4_2 \mathcal{N}_{2W^4} = 0\ ,\qquad
 W^4_3 \mathcal{N}_{3W^4} = 0\ ,
\end{equation}
and plugging in the structure constants \eqref{eq:qpStrConst} leads to the same expressions for $w^4$, 
$w^6$ and $c_{44}^{4}$ as those obtained in~\eqref{eq:opeP} and \eqref{eqn:gamma0} by associativity. Here 
$w^6$ is the eigenvalue of the zero mode of $W^6$ on $\Phi$.


\section{Structure constants of \texorpdfstring{$\hse$}{hse[mu]}}
\label{sec:hse}


The algebra $\hse$ is a subalgebra of $\hs{\mu}$ and the structure constants of the latter are known explicitly, 
see  \cite{Fradkin:1990qk}. 
We have rescaled the generators of this reference so that the first few 
commutation relations take the form
\begin{align}\label{hsecom}
 [L_m,W^s_n] &= \left((s-1)m-n\right)W^s_{m+n}\ ,\\
[W^4_m,W^4_n]&=-\frac{20}{\sqrt{7}} P^{44}_6(m,n) W^6_{m+n} 
+\frac{12}{\sqrt{5}} \left(\mu ^2-19\right) P^{44}_4(m,n)
   W^4_{m+n}\notag\\
&\quad{}+8 \left(\mu ^4-13 \mu ^2+36\right) P^{44}_2(m,n)  L_{m+n}
\ ,\displaybreak[3]\\
[W^4_m,W^6_n]&=-8 \sqrt{\frac{210}{143}} P^{46}_8(m,n) W^8_{m+n}+\frac{14}{\sqrt{5}} \left(\mu ^2-49\right)
    P^{46}_6(m,n) W^6_{m+n} \notag\\
&\quad{}-\frac{20}{\sqrt{7}} \left(\mu ^4-41 \mu ^2+400\right)  P^{46}_4(m,n) W^4_{m+n}\ ,\displaybreak[3]\\
[W^6_m,W^6_n]&=-252 \sqrt{\frac{5}{2431}}P^{66}_{10}(m,n) W^{10}_{m+n}
+28 \sqrt{\frac{6}{143}} \left(\mu ^2-115\right) P^{66}_8(m,n) W^8_{m+n}\notag\\
&\quad{}-\frac{40}{3 \sqrt{7}} \left(\mu ^2-88\right) \left(\mu ^2-37\right) P^{66}_6(m,n)W^6_{m+n}\notag\\
&\quad{}+\frac{14}{\sqrt{5}} (\mu^2 -49) (\mu^2 -25) (\mu^2 -16) P^{66}_4(m,n) W^4_{m+n}\notag\\
&\quad{}+12 (\mu^2 -25) (\mu^2 -16) (\mu^2 -9) (\mu^2 -4) P^{66}_2(m,n) L_{m+n} \ ,\\
[W^4_m,W^8_n]&=-20 \sqrt{\frac{6}{17}} P^{48}_{10}(m,n) W^{10}_{m+n} 
-\frac{72}{13 \sqrt{5}} \left(277-3 \mu ^2\right)P^{48}_8(m,n) W^8_{m+n}\notag \\
&\quad{}-40\sqrt{\frac{210}{143}} (\mu^2 -49) (\mu^2 -36) P^{48}_6(m,n) W^6_{m+n} \ ,
\end{align}
where $P^{ss'}_{s''}(m,n)$ are the universal polynomials containing the mode 
dependence of the structure constants in a commutator of quasiprimary fields of 
a CFT. They are given by
\begin{equation*}
P^{ss'}_{s''}(m,n) :=\sum_{r=0}^{s+s'-s''-1}\binom{s+m-1}{s+s'-s''-r-1}
\frac{ (-1)^r (s-s'+s'')_{(r)}(s''+m+n)_{(r)} }{r! (2s'')_{(r)}}\ ,
\end{equation*}
where we have introduced the Pochhammer symbols $x_{(r)}=\Gamma(x+r)/\Gamma(x)$.
When $m,n$ are restricted to the wedge, these universal polynomials 
are essentially the Clebsch-Gordan coefficients 
of $\mathfrak{sl}(2)$ \cite{Bowcock:1990ku}.
The proportionality factors between the generators $T^j_m$ of \cite{Fradkin:1990qk} and our generators $W^s_m$ are
explicitly
\begin{equation}
T^j_m = \sqrt{\frac{(j-m)!(j+m)!}{(2j)!}} W^{j+1}_m\ .
\end{equation}
%


\end{document}